\documentclass[journal]{vgtc}  
    \vgtccategory{Research}
    \vgtcpapertype{application/design study}
    
\usepackage{xargs}      
\usepackage{soul}       
\usepackage{xspace}     
\usepackage{xpunctuate} 
\usepackage{changepage}   

\newcommand{\etal}{{et~al\xperiod}\xspace}

\newcommandx{\guest}[3][1=]
    {\setulcolor{lightorange}{\ul{#1}} \textcolor{lightorange} 
    {[\textbf{#2:} #3]}}
\newcommandx{\shaq}[2][1=] 
    {\setulcolor{purple}{\ul{#1}} \textcolor{purple}   
    {[\textbf{Shakiba:} #2]}}  

\newcommand{\badge}[2]{\colorbox{#1}{\textcolor{white}{\textsc{#2}}}}

\newcommandx{\due}[1]{\badge{purple}{Due: #1}}  

\usepackage{graphicx} 
\graphicspath{{files/}{figs/}{Template/}{pictures/}{images/}{./}} 

\newcommand{\fignonvisual}{
\begin{figure}[htb]
    \centering
    \includegraphics[width=.5\textwidth]{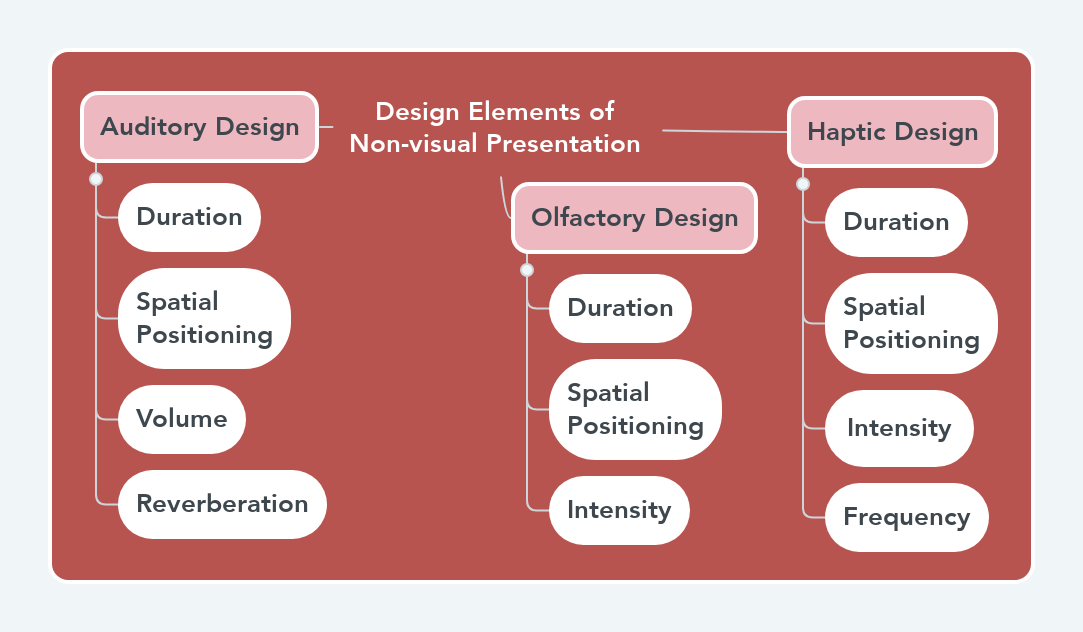}
    \caption{XR design elements for non-visual modalities}
    \label{fig:fignonvisual}
\end{figure}
}

\newcommand{\figAccuracySetting}{
\begin{figure}[htb]
    \centering
    \includegraphics[width=0.5\textwidth]{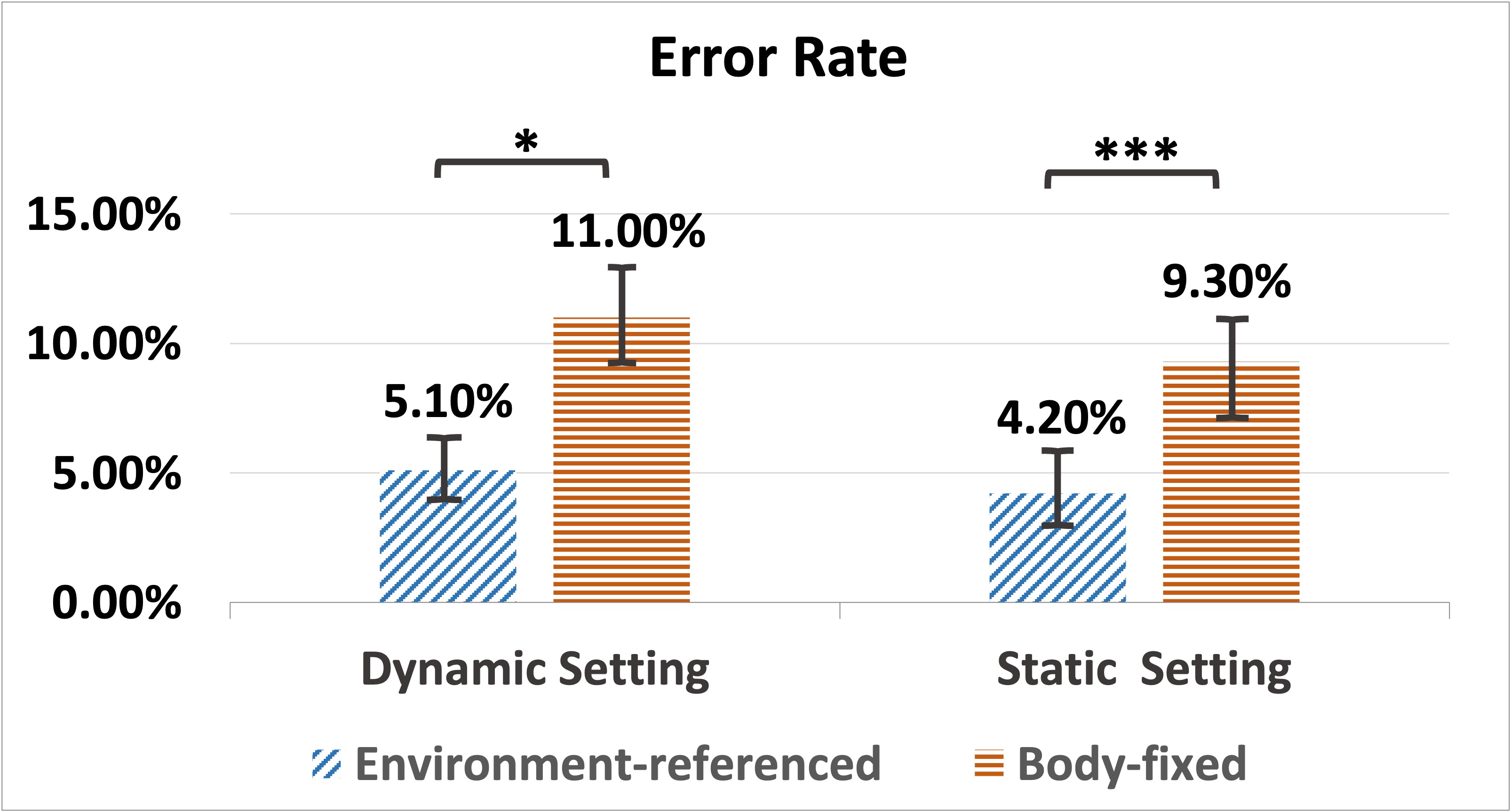}
    \caption{Accuracy within various RW settings}
    \label{fig:accuracy}
\end{figure}
}

\newcommand{\figNavTime}{
\begin{figure}[htb]
    \centering
    \includegraphics[width=0.5\textwidth]{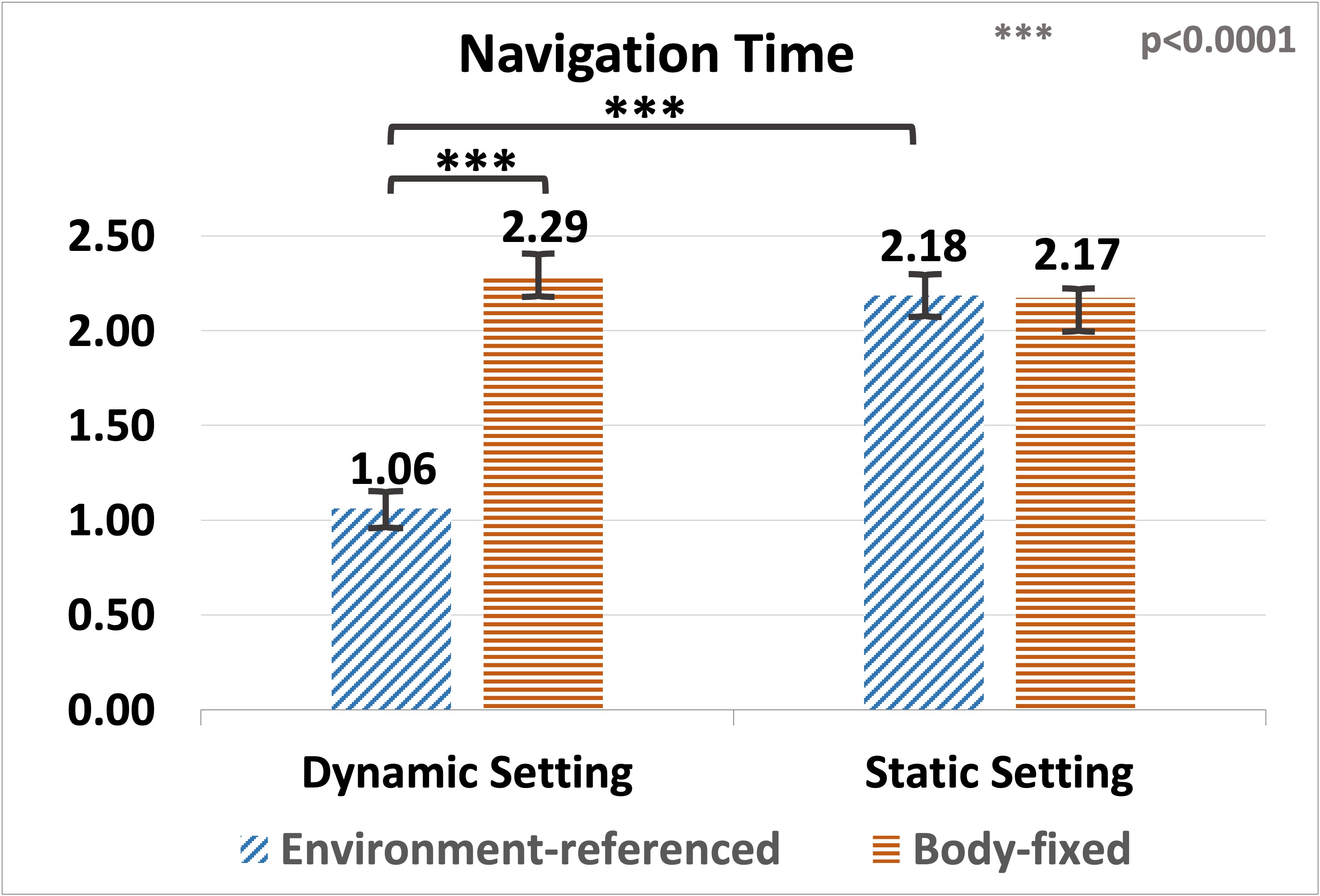}
    \caption{Time efficiency within various RW Settings}
    \label{fig:time}
\end{figure}
}

\newcommand{\figGazeContext}{
\begin{figure}[htb]
    \centering
    \includegraphics[width=0.5\textwidth]{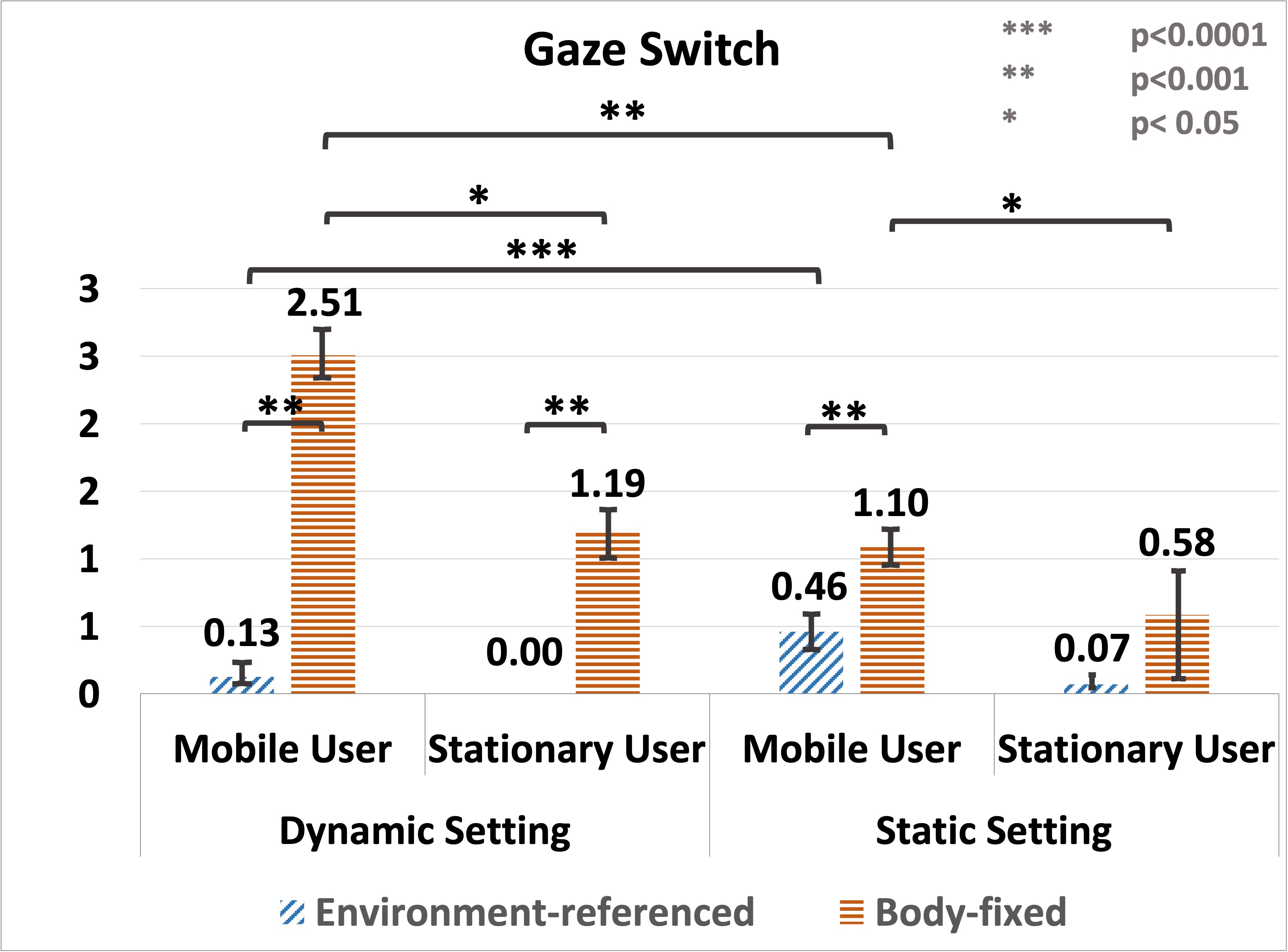}
    \caption{Gaze switch efficiency within various contexts}
    \label{fig:gazeContext}
\end{figure}
}

\newcommand{\figPreferred}{
\begin{figure}[htb]
    \centering
    \includegraphics[width=0.4\textwidth]{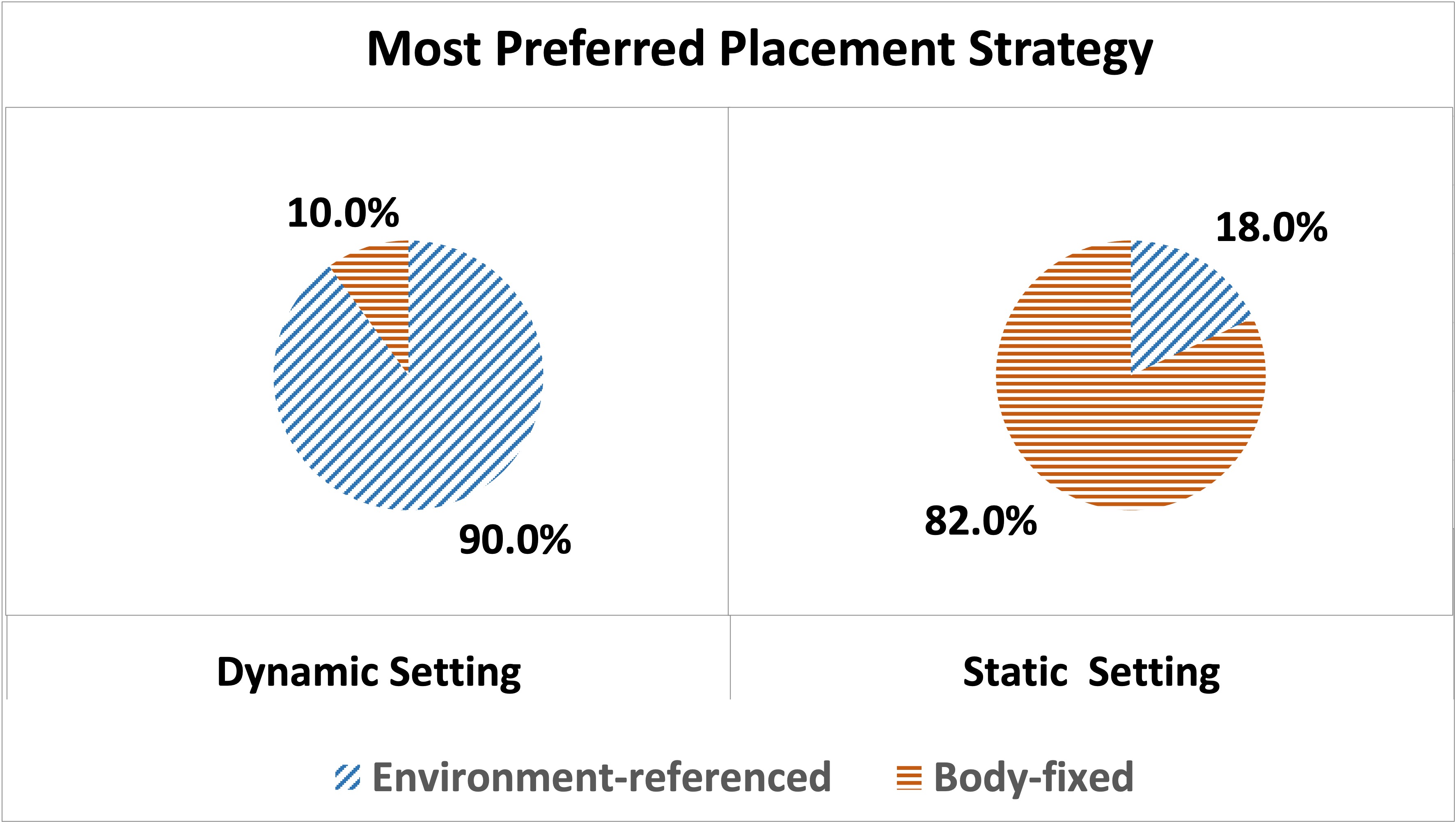}
    \caption{User preference on placement strategy in various RW settings}
    \label{fig:preferred}
\end{figure}
}

\newcommand{\figstaticGaze}{
\begin{figure}[htb]
    \centering
    \includegraphics[width=0.5\textwidth]{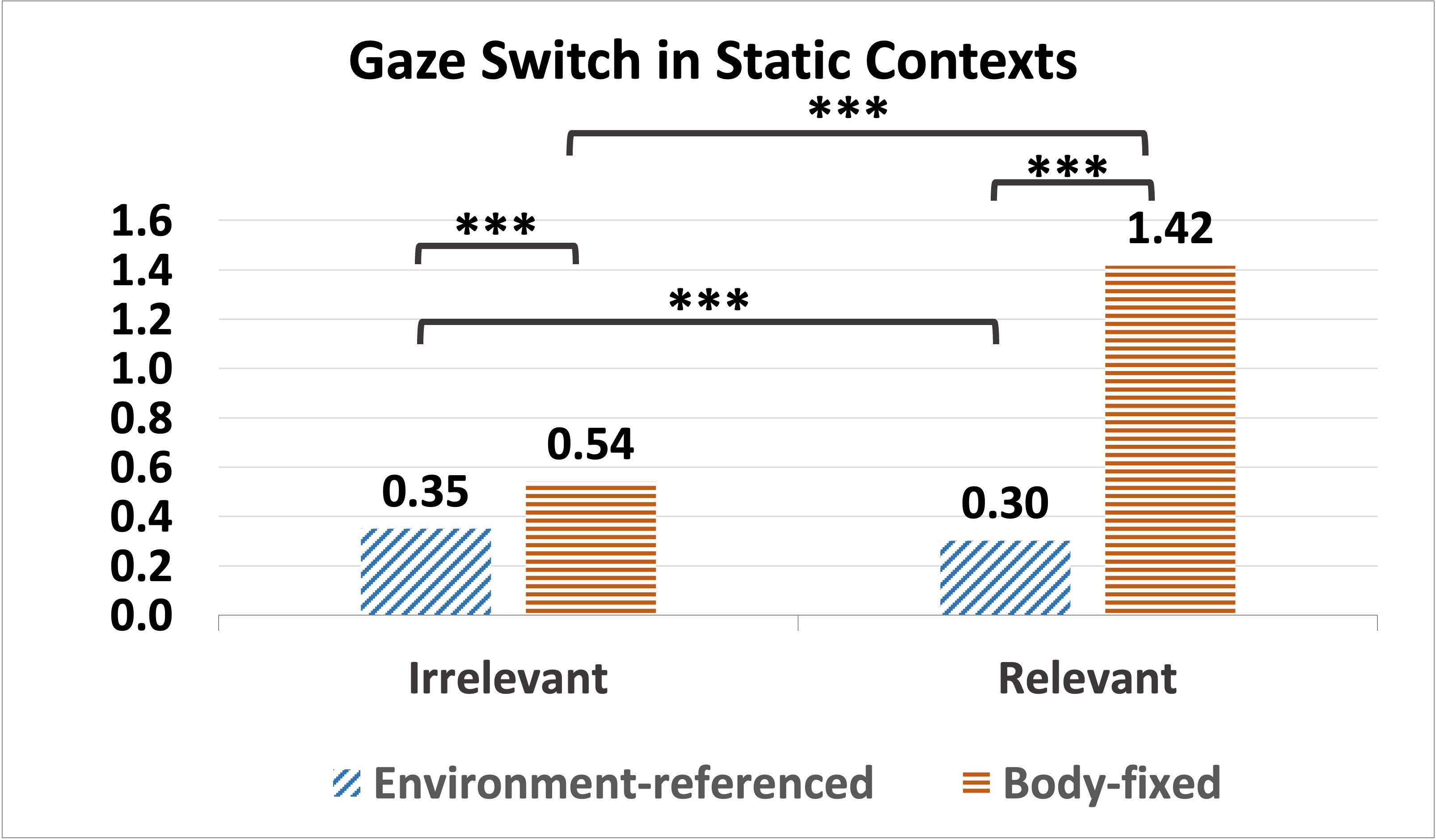}
    \caption{The influence of Info-focus Relevance on gaze switch efficiency within static contexts}
    \label{fig:gazeStatic}
\end{figure}
}

\newcommand{\figstaticTime}{
\begin{figure}[htb]
    \centering
    \includegraphics[width=0.5\textwidth]{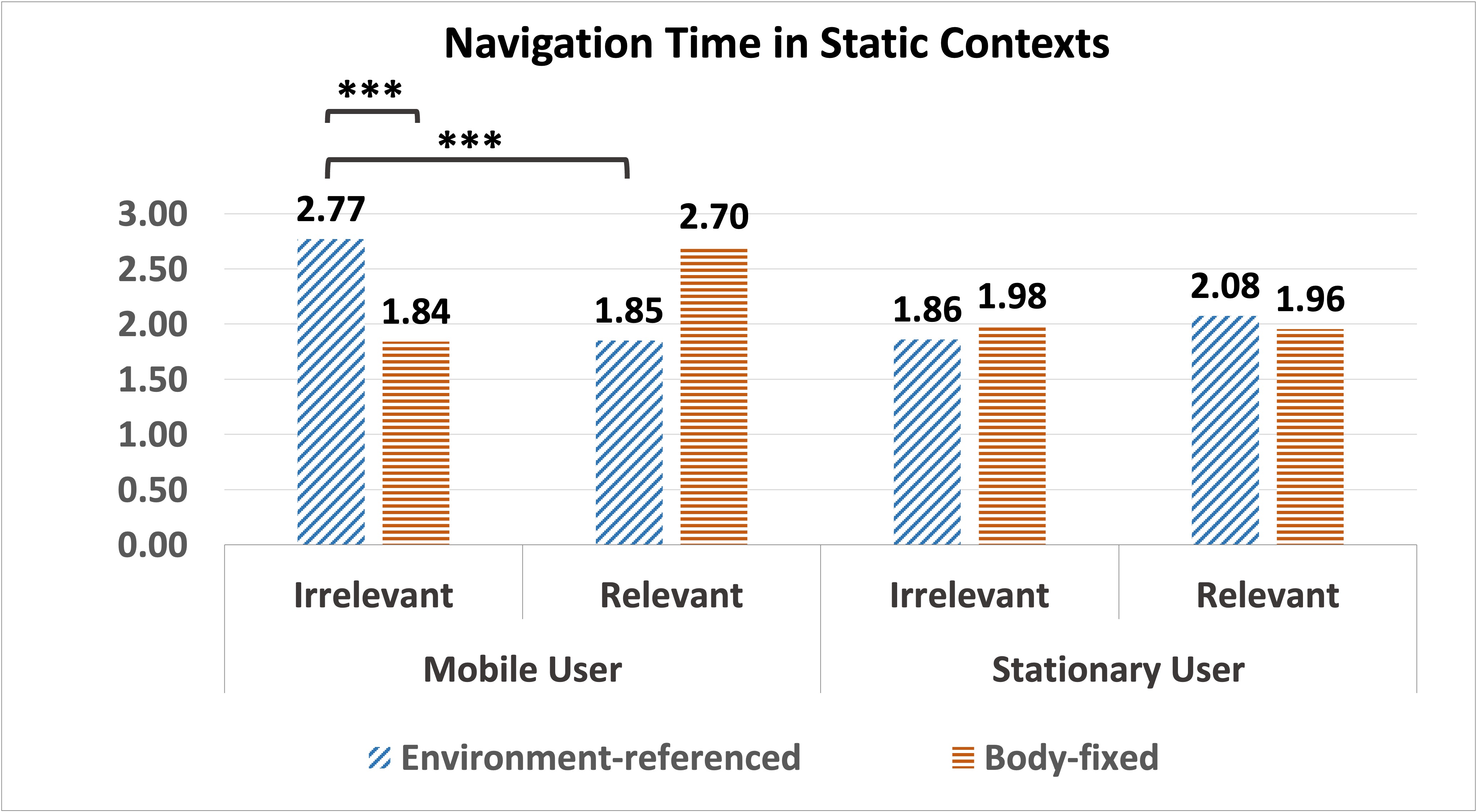}
    \caption{The influence of Info-focus Relevance on time efficiency within static contexts}
    \label{fig:timeStatic}
\end{figure}
}

\newcommand{\figExperimentDiagram}{
\begin{figure*}[t!]
    \centering
    \hspace*{-5em}
    \includegraphics[width=.8\linewidth]{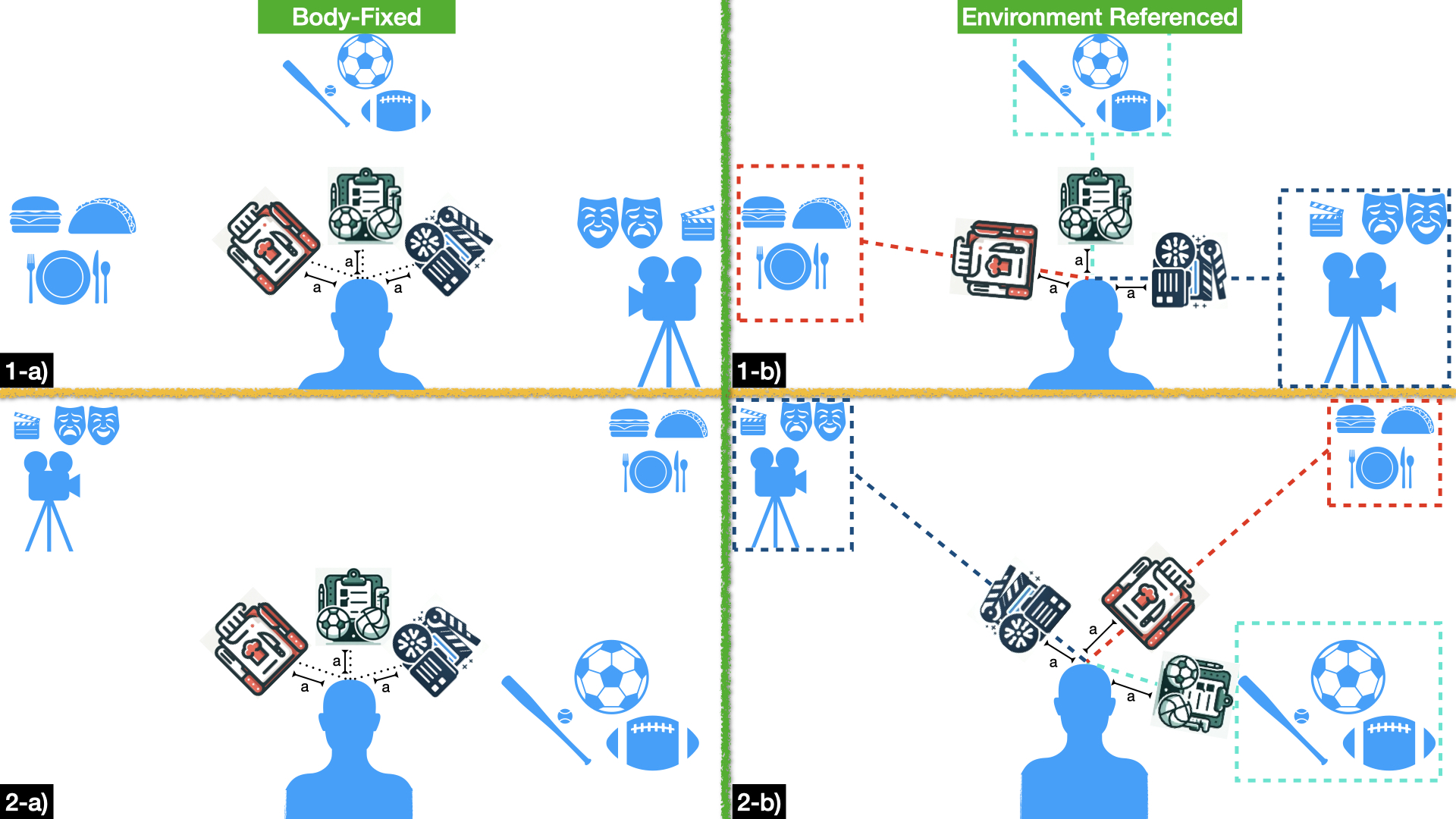}
    \caption{Illustration of our two placement strategies. 
    \textbf{(1-a, 2-a)}: Body-fixed placement: each XR object's pose remains consistent based on the orientation of the user, i.e., frame of reference.
    \textbf{(1-b, 2-b)}: Environment-referenced placement: each XR object's pose remains consistent based on the orientation of its intermediary relative to the frame of reference, maintaining a consistent distance from the user.
    \textbf{(1-a, 1-b)}: The user is at the session start point, positioned at an equal distance from the intermediaries.
    \textbf{(2-a, 2-b)}: The user has moved near the Sports category's intermediary.}
    \label{fig:ExperimentDiagram}
\end{figure*}
} 

\newcommand{\figsocialObj}{
\begin{figure}[htb]
    \centering
    \includegraphics[width=0.45\textwidth]{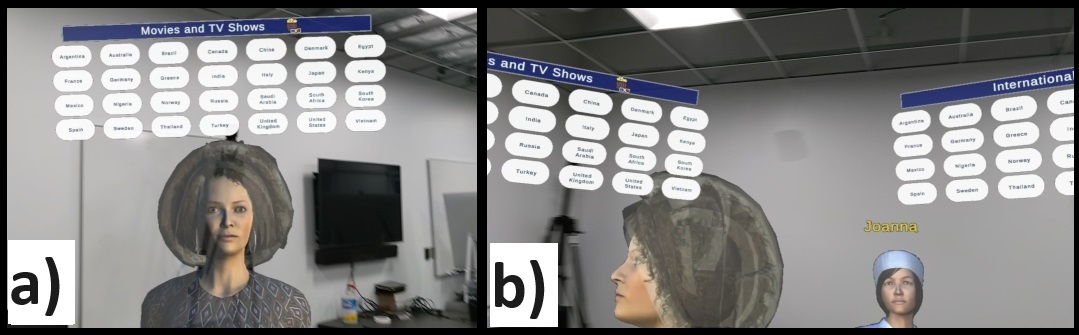}
    \caption{Environment-referenced placement in dynamic contexts: a) The category panel relevant to each trivia host is displayed above their head. b) the panels move along with their associated trivia host.}
    \label{fig:socialObj}
\end{figure}
}

\newcommand{\figsocialUser}{
\begin{figure}[htb]
    \centering
    \includegraphics[width=0.45\textwidth]{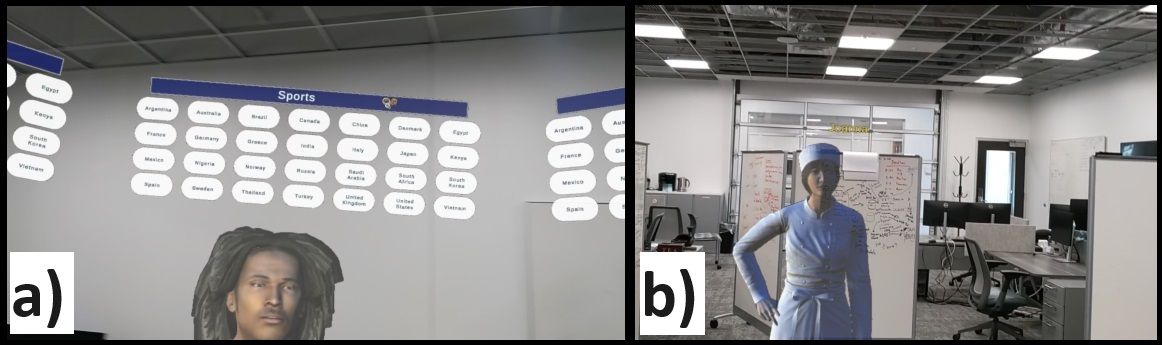}
    \caption{Body-fixed placement in dynamic contexts: a) All three category panels in front of the user's body. b) No AR content is placed near the trivia hosts, not directly in front of the user.}
    \label{fig:socialUser}
\end{figure}
}

\newcommand{\figdocument}{
\begin{figure}[htb]
    \centering
    \includegraphics[width=0.45\textwidth]{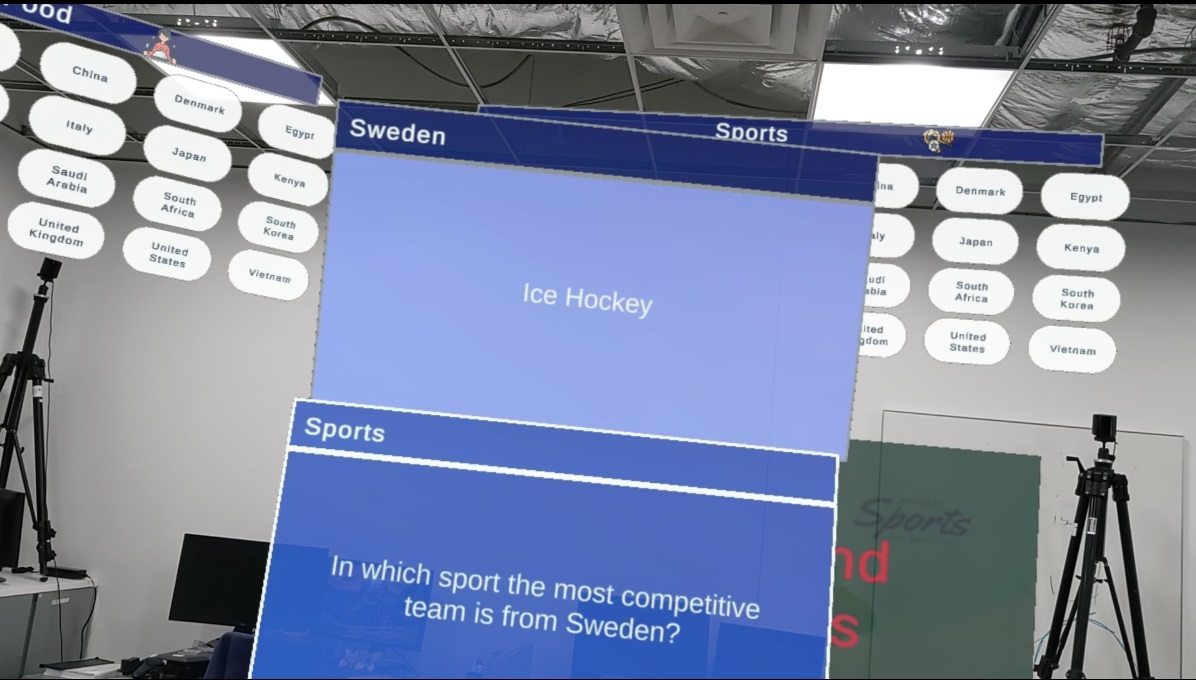}
    \caption{Each document displays the answer to a trivia question}
    \label{fig:document}
\end{figure}
}

\newcommand{\figquestion}{
\begin{figure}[htb]
    \centering
    \includegraphics[width=0.5\textwidth]{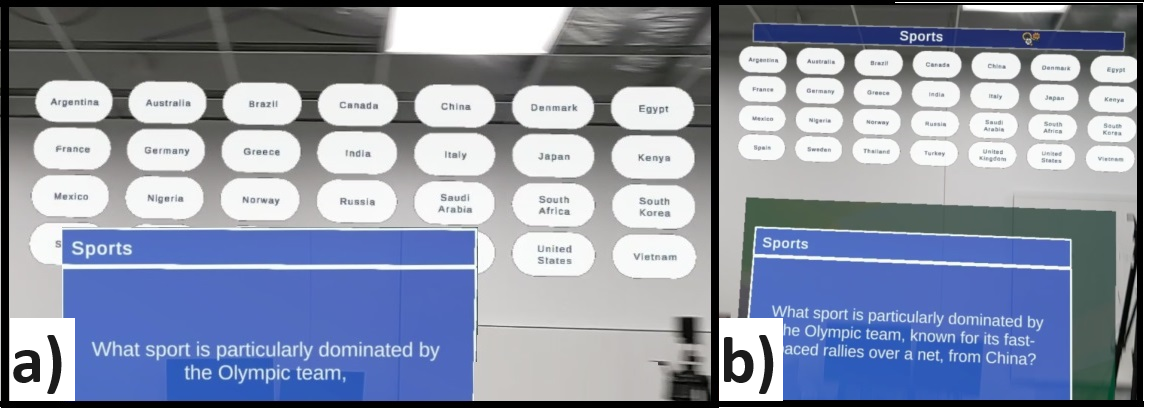}
    \caption{ a) The questions are presented word-by-word. b) The header for each category panel reappears after the question is asked.}
    \label{fig:question}
\end{figure}
}

\usepackage{comment}
\usepackage{mathptmx}  
    \ifpdf
  \pdfoutput=1\relax                   
  \usepackage{graphicx}                
  \DeclareGraphicsExtensions{.pdf,.png,.jpg,.jpeg} 
\else
  \usepackage{graphicx}                
  \DeclareGraphicsExtensions{.eps}     
\fi%

\usepackage{microtype}                 
\PassOptionsToPackage{warn}{textcomp}  
\usepackage{textcomp}                  
\usepackage{mathptmx}                  
\usepackage{times}                     
\usepackage{cite}                      
\usepackage{tabu}                      
\usepackage{booktabs}                  

    \title
    {Towards Context-Aware Adaptation in Extended Reality: A Design Space for XR Interfaces and an Adaptive Placement Strategy}

\author{
        \authororcid{Shakiba Davari}{0000-0003-3128-1979}
            \thanks{e-mail: sdavari@vt.edu}
        \and \authororcid{Doug. A. Bowman}{0000-0003-0491-5067}
            \thanks{e-mail: dbowman@vt.edu}
}
    \affiliation{\scriptsize Center for Human-Computer Interaction \\ Department of Computer Science, Virginia Tech, Blacksburg, VA, USA}

    \abstract{%
        By converting the entire 3D space around the user into a screen, Extended Reality (XR) can ameliorate traditional displays' space limitations and facilitate the consumption of multiple pieces of information at a time.
        However, if designed inappropriately, these XR interfaces can overwhelm the user and complicate information access. 
        In this work, we explored the design dimensions that can be adapted to enable suitable presentation and interaction within an XR interface.
        To investigate a specific use case of context-aware adaptations within our proposed design space, we concentrated on the spatial layout of the XR content and investigated non-adaptive and adaptive placement strategies.
        In this paper, we 
        \textbf{(1)} present a comprehensive design space for XR interfaces, 
        \textbf{(2)} propose \textit{Environment-referenced}, an adaptive placement strategy that uses a relevant intermediary from the environment within a Hybrid Frame of Reference (FoR) for each XR object, and
        \textbf{(3)} evaluate the effectiveness of this adaptive placement strategy and a non-adaptive \textit{Body-Fixed} placement strategy in four contextual scenarios varying in terms of social setting and user mobility in the environment.
        The performance of these placement strategies from our within-subjects user study emphasized the importance of intermediaries' relevance to the user's focus.
        These findings underscore the importance of context-aware interfaces, indicating that the appropriate use of an adaptive content placement strategy in a context can significantly improve task efficiency, accuracy, and usability.
    }
    \teaser{
        \centering
        \includegraphics[width=.99\linewidth]{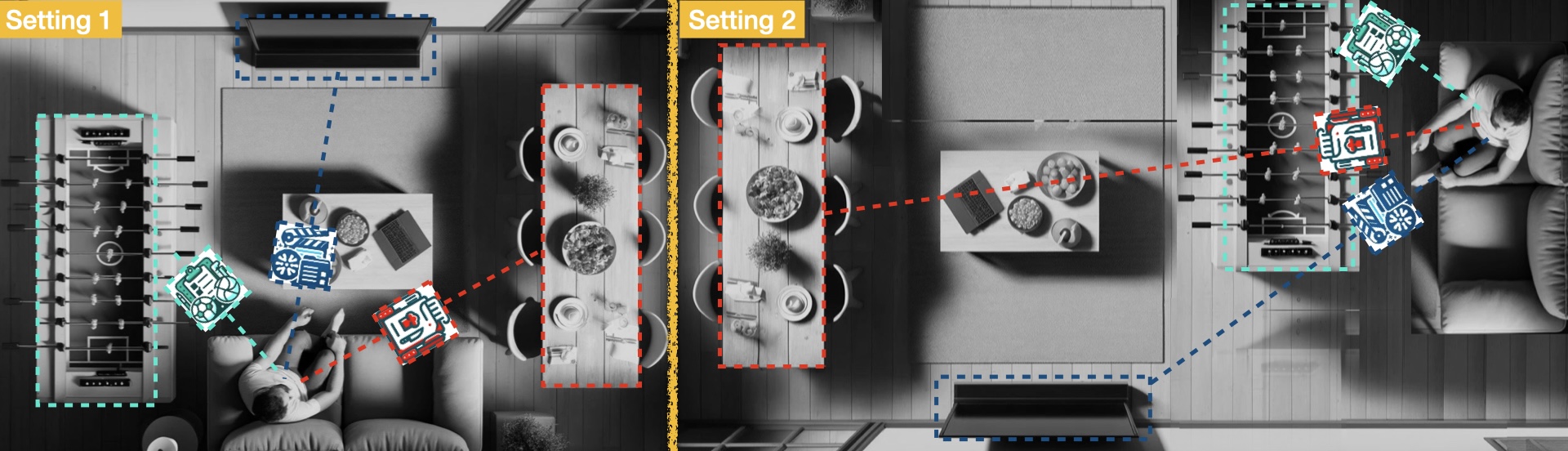}
        \caption{Environment-referenced placement strategy: to adaptively place an XR object within varying contexts, this strategy identifies a relevant intermediary from the environment to be utilized within the object's frame of reference.}
        \label{fig:teaser}
    } 
    \renewcommand{\manuscriptnotetxt}{}
    
\begin{document}
\firstsection{Introduction}
\maketitle
Extended Reality (XR) is envisioned to become a primary modality for information access, providing continuous, effortless, and hands-free access to multiple pieces of information through a glance, anywhere, and at any time \cite{davari2020occlusion, lu2020glanceable,lu2021evaluating, feiner2002augmented}.
For instance, consider a conversation about scheduling a hike on a sunny day. Using a mobile phone involves multiple steps to 1) retrieve the device, 2) unlock the screen, 3) locate the calendar and weather apps, and 4) frequently switch between them. Furthermore, this process diverts the user's attention away from their conversation partner and social cues.
In this scenario, Augmented Reality (AR) offers simultaneous access to both apps, eliminating the need for frequent switching, simplifying the process of locating the apps. Moreover, placing the apps near the interlocutor's face can enhance social awareness and engagement in the conversation \cite{davari2022validating}.
\par
However, these XR interfaces may also increase intrusiveness, environmental occlusion, and information overload, posing challenges to the user's perception and awareness of their surroundings.
XR's limitless possibilities for providing information pose a unique challenge for designers, who must create interfaces that seamlessly integrate into users' environments while maintaining usability and clarity.
For instance, when overlaying content onto a busy environment, maintaining content visibility while avoiding intrusions to the user's interactions with other real-world or physical objects can be challenging.
Failure to design a suitable interface can diminish performance (e.g., illegible presentiation of instructions in a virtual training simulation), or even environment awareness and safety (e.g., occlusion of a car by virtual navigation cues when crossing the street).
\par
Identifying the XR interface's design elements that can adapt is of utmost significance in designing efficient yet non-intrusive XR interfaces.
Numerous adaptation strategies can be deployed within an interface to utilize various contextual information.
Consider a scenario in which Miles and Alex meet at the library to review the course material together and prepare for an exam. Overlaying documents on Alex's face can disrupt Miles' social interactions with Alex. In this context, an optimal interface tracks Alex's face and adapts to prevent its occlusion.
\par
However, the mere presence of such adaptations does not necessarily ensure optimal effectiveness.
The efficiency of specific adaptive and non-adaptive interfaces is contingent upon the specific context of their usage \cite{davari2022dc, grubert2016towards, lindlbauer2019context}.
For instance, in the scenario above, after reviewing the material, Miles and Alex decide to study the next chapter separately.
While the overall scenario remains unchanged, preventing occlusion of Alex's face can now lead to suboptimal placement, inconsistency, and distraction, as Miles is no longer interacting with him.
Conversely, a non-adaptive placement may occlude Alex's face, yet consistantly offer efficient access and minimize distractions.
Designing an effective XR interface requires identifying the relevant XR content and their optimal presentation and interaction within various contexts \cite{davari2022dc}.
\par
Towards context-aware adaptation in XR, we explore three fundamental Research Questions (\textbf{RQs}) investigating the XR design space (\textbf{RQ1}), utilizing this design space to propose an adaptive spatial placement strategy (\textbf{RQ2}), and exploring context-awareness through empirical evaluation of a non-adaptive and an adaptive placement strategy in various contexts (\textbf{RQ3}).\\
\\
\noindent
\textbf{\textit{RQ1:}} What are the \textbf{design elements} of an XR interface?\\
An XR interface is composed of a set of interactable XR objects that present information in various modalities.
Adapting the design elements of these XR objects enables context-aware interfaces that optimize the provided XR content and its presentation within a specific context. 
In \Cref{se:designSpace} we examine potential XR adaptations and outline a design space for XR.
\\
\\
\noindent
\textbf{\textit{RQ2:}} How can we adapt the \textbf{spatial placement} of XR objects utilizing context?\\
The inappropriate spatial layout of XR objects can lead to distractions and occlusions, overwhelm the user, and hinder their interactions with the environment.
\Cref{se:caPlacement} proposes an adaptive placement strategy utilizing an intermediary object from the environment within a hybrid frame of reference.
This placement strategy leverages contextual information to maintain efficiency and unobtrusiveness within context-switching scenarios.
\\
\\
\noindent
    \textbf{\textit{RQ3:}} Does adaptive XR placement \textit{always} enhance performance?\\
    While an adaptive placement strategy adjusts to specific contextual information, relying solely on one or a few context components may be insufficient for optimal performance.
    The people, objects, events, and elements within the environment can affect the user's behavior, information access needs, and interests.
    Similarly, one's physical, cognitive, and emotional states influence perception and interactions within the environment, changing perspective, spatial awareness, directional perception, and distance from different objects.
    Varying the setting and the user state can influence the effectiveness and performance of an adaptive placement strategy within a scenario. 
    \Cref{se:experiment3} designs an experiment to explore the nuanced interplay between these contextual components and the effectiveness of placement strategy.
    This experimental design evaluates the accuracy, efficiency, and usability of our proposed adaptive placement strategy and a traditional body-fixed placement across four distinct contexts.
    \Cref{se:results} presents the results from our user study within \textit{Dynamic Mobile}, \textit{Static Mobile}, \textit{Dynamic Stationary}, and \textit{Static Stationary} contexts.
\section{Related Work}
Previous work explored various innovative applications and design dimensions of XR to spatially place 2D \& 3D objects and transition between them, visualize hierarchies, and provide persistent and portable presentation of the personal information \cite{di2003arwin, feiner1993windows, lu2021evaluating, dunston2008identification, langlotz2014next}.
Morrison et al. highlighted unique design elements within AR for enhancing accessibility for visually impaired children \cite{ed2021socialsensemaking}.
These studies underscore the broad design space of XR interfaces and the versatile and transformative applications that XR enables across various contexts. This work, investigates previous work and the design elements they utilized, providing a comprehensive XR design space.
\par
XR offers the potential to enable efficient information access, reduce cognitive load, and enhance user convenience compared to traditional methods such as mobile phones \cite{valimont2002effectiveness, davari2022validating, matthews2007designing, davari2020occlusion, lu2020glanceable}. 
However, intrusive XR interfaces may result in challenges such as information overload and occlusion of important cues within the environment \cite{krevelen2010survey, bengler2006augmented}, increase cognitive load and discomfort, and reduce the user’s situational awareness and performance \cite{endsley1988design, grubert2010extended, stoltz2017augmented}.
Various approaches for intuitive and seamless integration of XR content into the environment have been extensively explored. %
For instance, to enhance efficiency and minimize intrusiveness, numerous designs adapt the XR content's availability, transparency, placement, and Level of Detail (LoD) \cite{diverdi2004level, bell2001view, pfeiffer2008evaluation}, as well as spatial layout and size \cite{ens2014personal, cheng2021semanticadapt}.
Lages \& Bowman highlighted the significance of adapting the AR content placement strategy to avoid occlusions and accommodate activities like walking \cite{lages2019walking}.
For adaptations to the XR content placement, the concept of the frame of reference, also referred to as fixation was introduced \cite{feiner1993windows}. 
\par
User-triggered adaptation through gaze, hand, and head-based inputs such as finger taps and handheld controllers are extensively explored for adjustments to the transparency, LoD, and spatial layout of XR content \cite{ens2014personal, lages2019adjustable, rivu2020stare, lu2021sui, ken2021, pfeuffer2021artention}.
In AR, for instance, many applications were designed to prioritize the real-world \cite{davari2020occlusion} by initially keeping the XR content hidden, in the peripheral, or at a lower LoD, and granting access to them through explicit interactions \cite{piening2021looking, matthews2006designing, pfeiffer2008evaluation}.
However, AR's definition emphasizes the \textit{integration of the digital content into the real world} \cite{azuma1997survey}, underscoring the significance of context awareness.
\par
While user-triggered adaptations offer control and predictability, they increase the user's physical and mental workload of deciding when, what, and how to apply the adaptations \cite{roy2019automation}.
Automatic XR adaptation can enhance efficiency and reduce workload compared to the user-triggered ones \cite{davari2020occlusion}. 
Numerous studies suggest rule-based approaches for XR adaptations.
The significance of such rule-based adaptations in meeting the XR task requirements within various applications such as driving and conversation have been highlighted \cite{bengler2006augmented, davari2022validating}. 
To prevent occlusion issues, Ens et al. suggested a rule-based adaptive design to exclusively place the XR objects on empty surfaces \cite{ens2015spatial}.
Constraints, explicitly imposed by the users, were utilized as guidelines to group related XR objects together and prevent their occlusion within a rule-based view management \cite{bell2001view}.
Such rule-based adaptive approaches are highly tailored to specific use cases and applications.
Even within the same application or use case, slight contextual deviations can cause a rule to fail, making it suitable only within unchanging contexts.
This work proposes an adaptive placement strategy, applicable within changing contexts, to extract and utilize contextual information from the environment and user state to spatially place the XR content.
\par
Context refers to the external components that influence or relate to the user's interactions with the interface \cite{dey2001conceptual}.
In recent years, context-aware XR has become a focal point of research, promising the potential for ``ubiquitous" and ``pervasive" computing through AR \cite{weiser1999computer, grubert2016towards}.
Contextual aspects such as user preferences, cognitive load, device profiles, task environment, semantic changes, and task-specific security parameters have been utilized for adaptations to the XR content's appearance, LoD, frame of reference, and spatial layout \cite{lindlbauer2019context, cheng2021semanticadapt, lee2008visualization}.
\section{Design Space of XR Interfaces} \label{se:designSpace}

\begin{figure*}[t!]
    \vspace*{-2em}
    \centering
    \hspace*{-3em}
    \includegraphics[width=1.1\linewidth]{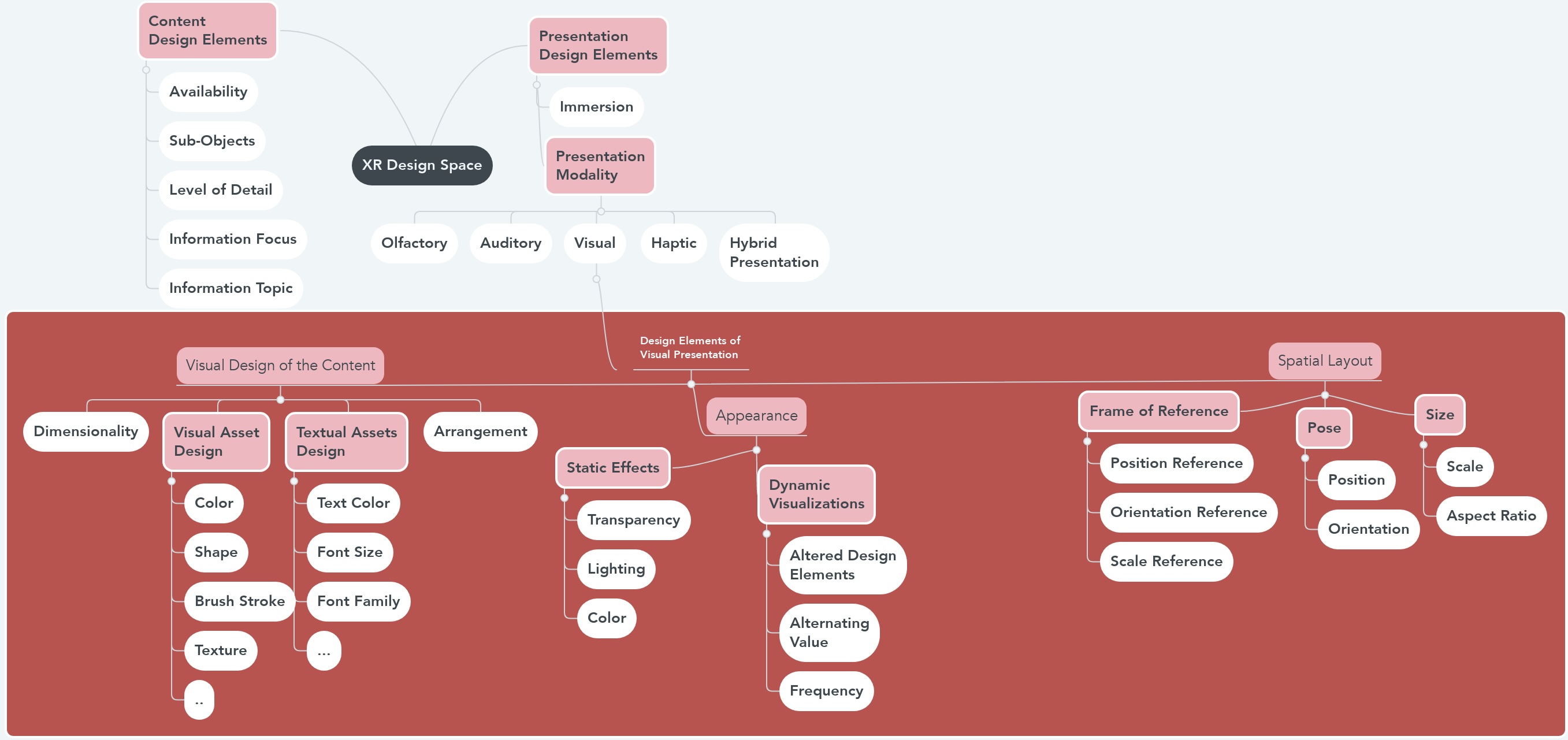}
    \caption{Design Space for XR interfaces}
    \label{fig:ds}
    \vspace*{-1.7em}
\end{figure*}
An intelligent XR interface dynamically adapts the content, presentation, and interaction of each XR object to contextual changes. Prior research extensively investigates design elements and adaptation techniques for XR interfaces \cite{bowman2001interaction, azuma1997survey, grubert2021MRinteraction, schmalstieg2017augmented}. We draw from the existing literature to identify these design elements and propose a comprehensive design space for XR interfaces (see \Cref{fig:ds}).
\par
We regard an XR object as the fundamental component of any XR interface, irrespective of its modality for presenting digital information or its potential for interactivity. 
Each XR object may comprise multiple other sub-objects and can provide applications, auditory notifications, 3D models, 2D images, system recommendations, menus, and the like.
We divide the design elements of XR objects into two high-level categories: content design and presentation design (\Cref{fig:ds}).
\subsection{Content Design} \label{subse:contentadapt}
    Previous research explored adapting the \textbf{availability} of XR content to provide the right information at the right time, enhancing performance and preventing potential obstructions and distractions in various contexts \cite{lindlbauer2019context, bengler2006augmented, davari2022validating}.
    Effectively adapting in-use XR objects' \textbf{availability} allows the creation of interfaces that ``close", ``open", or ``minimize" apps tailored to the user.
    Within the design of an XR interface, the availability of each XR object can be modified either manually by the user or through context-aware adaptations, or be immutable (e.g., a constantly ``opened" system-level XR object).
    For instance, in educational XR aimed at facilitating student learning experiences, the availability of specific XR objects may be contingent upon the student's location, fulfillment of predetermined criteria, or their class and exam schedule.
    \par
    Each XR object is dedicated to a specific subject inherent to its essence and remains immutable.
    For instance, one XR object functions as a weather app while another provides recommendations or notifications to the user.
    However, within an interface, adaptations can also be made to the content of an XR object, 
    varying its \textbf{sub-objects}, \textbf{Level of Detail (LoD)}, \textbf{information focus}, and \textbf{topic} \cite{di2003arwin, diverdi2004level, lindlbauer2019context}.
    For instance, 
    the content within the weather app could display data focused on either the hourly forecast or the 10-day forecast, exclusively feature precipitation forecasts, or offer minimal LoD, showcasing solely visualizations of the current temperature and conditions.
    Adapting the content within an XR object allows for the creation of a personalized interface, offering only the intended content.
    Imagine an XR object that provides recommendations to the user. Context-aware adaptations to this XR object's content facilitate the recommendation of various apps and information tailored to the user's needs, thereby enhancing their performance. 
    The objective of an effective XR interface is to present pertinent information in a timely and seamlessly integrated manner, ensuring it remains intuitive, easily accessible, understandable, and unobtrusive to the user.
    This underscores the pivotal role of content presentation in the design process of XR objects.
\subsection{Presentation Design} 
    The presentation design of an XR object involves its immersion and the modality in which it is presented (refer to \Cref{fig:ds}). 
    The \textbf{immersion} design element of an XR object determines whether it is immersive or non-immersive.
    A non-immersive XR object allows simultaneous interaction with other XR objects. However, an immersive XR object restricts user engagement solely to itself and its sub-objects, limiting interaction with external XR objects.

\vspace*{3pt}

\noindent
    \textbf{\textit{Presentation Modality}} \\  
    Skarbez \etal propose that all technology-mediated experiences fall under the term mixed reality, suggesting that an XR object can present information through various modalities \cite{skarbez2021revisiting}. The efficacy of an XR interface and the information it provides also relies on its presentation modality.
    In modern XR interfaces, XR objects are predominantly showcased as 2D and 3D overlays within the XR environment, utilizing the \textit{visual modality}. However, XR objects can possess the capacity to convey content through \textit{audio, haptic, and olfactory modalities}, as well as through \textit{hybrid presentations}.
    \subsubsection{Visual Presentation}
        Extensive research in the literature has investigated the visual design of XR interfaces and their design elements. 
        We propose a taxonomy for the visual design elements of XR objects, categorizing them into the visual design of the content of an XR object, and the XR object's appearance, and spatial layout within the environment (\Cref{fig:ds}).

\vspace*{3pt}

\noindent
        \textbf{\textit{Visual Design of the Content}}\\
        An XR object's visual content involves a structured arrangement of various 2D or 3D visual and textual assets, impacting the clarity, readability, and comprehensibility of the XR object. The \textbf{dimensionality} design element indicates whether the XR object employs 2D or 3D visualizations to present its content, while the \textbf{arrangement} design element determines how these assets are organized and integrated to create a coherent visualization.
        \par
        \textbf{\textit{Visual Asset Design:}}
        Design elements such as \textbf{texture}, \textbf{brushstroke}, \textbf{size}, and \textbf{color} are instrumental in designing 2D or 3D images, models, lines, and shapes \cite{laviola2022minimal}.
        These visual assets represent either the entirety or specific aspects of an XR object's content.
        \par
        \textbf{\textit{Textual Asset Design:}}
        Design elements pertinent to the typography of the text influence the legibility, readability, and aesthetic appeal of textual assets, such as text boxes \cite{baines2005type}. 
        These typography features encompass various design elements such as \textbf{typeface selection}, \textbf{text color}, \textbf{font and point size}, \textbf{line length}, \textbf{line and letter spacing}.
        Typography design elements can serve readability purposes and creative endeavors, such as conveying specific messages and establishing hierarchy, tone, and visual rhythm.
        For instance, context-aware adaptations to typography enable XR to accommodate low-vision or dyslexic users.

\vspace*{3pt}

\noindent
        \textbf{\textit{Appearance}}\\
        A multitude of static and dynamic design elements collectively influence the realism, perceptibility, and noticeability of an XR object's appearance.
        Previous work proposed guidelines for limiting unnecessary or non-conforming dynamic visuals and effectively using visual cues, color, and animation to enhance performance and prevent potential obstructions and distractions \cite{schmalstieg2017augmented, bengler2006augmented}. 
        These elements contribute to the natural integration and alignment of visual XR objects within the XR environment.
        \par
        \textbf{\textit{Static Effects:}}
        Various visual effects can be applied to manipulate the \textbf{transparency}, \textbf{lighting}, and \textbf{colors} of the 2D or 3D visualization of an XR object.
        Some common static visual effects include night/day mode color themes and adaptations to color saturation, warmth, tint, brightness, and contrast. 
        These adaptations significantly impact the opacity, legibility, and intuitiveness of the XR interface within its context. 
        For example, adjusting an XR object's transparency and colors enables its seamless integration into the XR environment \cite{bell2001view}.
        Similarly, aligning the shadow and brightness of an XR object with the environmental lighting conditions ensures its noticeability without overwhelming brightness, enhancing the natural appearance of the scene.
        \par
        \textbf{\textit{Dynamic Visualizations:}}
        Adapting at least one design element of an XR object at a specified frequency results in dynamic 2D or 3D animated visualizations within the XR object. For example, an animated XR object that dynamically moves at speed $v$ visualizes pose adaptations to its position design element at the frequency of $v$. While dynamic visualizations do not necessarily follow a periodic pattern, their design elements can consist of the \textbf{altered design elements} and their \textbf{alternating values} and \textbf{alteration frequencies}. Such dynamic visualizations can enhance the overall realism, noticability, and interactivity of the XR interface, and the user's awareness by highlighting relevant information or events within specific XR objects.
        Consider a dynamic visualization that alters the content compositing of an XR object by altering its line colors between green and transparent at frequency $f$. 
        Such blinking outlines can provide notifications about new mission assignments in an immersive VR game or indicate the occlusion of an important object in a context-aware AR \cite{davari2020occlusion}. 

\vspace*{3pt}

\noindent
        \textbf{\textit{Spatial Layout}} \\ 
        The spatial integration of a visual XR object plays a crucial role in 
        the interface's efficiency and the user's engagement and safety \cite{ens2014personal, lages2019adjustable}.
        The proper placement and alignment of XR objects can facilitate and speed up access to XR content, prevent occlusions, and influence the user's perception and awareness of the virtual and real-world environments. 
        To ensure that an XR object is displayed in a non-intrusive, realistic, and easily accessible manner within the user's field of view, it is essential to give careful consideration to the design of the frame of reference, the object's pose, and size.
        \par
        \textbf{\textit{Frame of Reference (FoR):}}
        A coordinate system is used as the FoR to determine the XR object's position, orientation, and scale in relation to the environment \cite{feiner1993windows, billinghurst1999wearable, bowman2001interaction, jose2016comparative, ens2014personal, billinghurst2015survey}.
        The literature has identified various FoRs for registering XR objects within the environment as world-fixed, object-fixed, head-fixed, body-fixed, or device-fixed \cite{bowman2001interaction}.
        \textit{World-fixed FoR }
        registers the XR object in a fixed position relative to stationary real-world locations and objects.
        \textit{Object-fixed FoR }
        is defined relative to another XR object or real-world object (including input devices), so that the content moves along with that object.
        \textit{Head-fixed FoR }
        registers XR objects relative to the virtual camera. This FoR is also referred to as display-fixed, Heads-Up-Display (HUD), or fixed display.
        \textit{Body-fixed FoR } 
        registers the XR objects relative to the user's body.
        World-fixed, display-fixed, and body-fixed settings have been commonly used as the FoR for XR objects in AR interfaces \cite{feiner1993windows}. 
        However, in all these scenarios, the coordinate system of the world or another object is uniformly employed as the FoR for position, orientation, and scale.
        For example, by utilizing $ref$ as the reference object for $O1$, any alterations to its position, orientation, and scale would apply to $O1$.
        \par
        While previous research studies commonly apply a \textbf{Unified FoR}, we propose that an XR object's FoR can be \textbf{Hybrid} and adhere to a distinct \textbf{position reference}, \textbf{orientation reference}, and \textbf{scale reference}. 
        Recent research by Manakhov \etal touches on the concept of hybrid FoRs, specifically discussing the \textit{HeadDelay} and \textit{Path} FoRs in the context of walking scenarios \cite{pavel2024FoR}.
        For instance, consider $N$ as an XR object positioned on the north side of the user and displaying a pointer towards the north direction in an AR navigation interface. As its FoR, this XR object can utilize a custom coordinate system that employs the user as its $reference_{position}$ and scale reference and the world as its orientation reference. 
        \begin{displaymath}
                N_{FoR} = (ref_{position}: user, \:\:
                            ref_{scale}: user, \:\:
                            ref_{orientation}: world )
        \end{displaymath}
        This FoR enables the XR object to be consistently positioned on the north side of the user and point towards the north regardless of the user's orientation while simultaneously following the user's position and maintaining a fixed distance from the user in mobile contexts.
        \par
        \textbf{\textit{Pose:}}
        The pose of an XR object in 3D space is specified through its \textbf{position} and its \textbf{orientation} relative to its FoR.
        Cartesian $(x,\: y,\: z)$ or spherical $(r,\: \theta,\: \phi)$ coordinates can denote the \textbf{position}, while the facing or pointing \textbf{orientation} can be indicated through Euler angles, quaternions, or rotation matrices.
        \par
        \textbf{\textit{Size:}}
        The \textbf{scale} of an XR object's size relative to its FoR, along with its \textbf{aspect ratio}, represents its size within the environment.
        Scale can be depicted by a singular value, such as a uniform scale factor, or by distinct scale factors for each dimension.
        The size of the XR object, including its length, width, and height within the environment, indicates the space it occupies, contributes to a sense of depth, and affects the realism and occlusion of the XR environment.
        Maintaining the correct aspect ratio can be essential in specific XR objects, as deviations may lead to distortion or misalignment, detrimentally affecting its integration within the environment and the user experience. 
    \subsubsection{Non-Visual Presentation}
        XR objects can be presented through \textit{audio, haptic, and olfactory modalities}, as well as through \textit{hybrid presentations}. As XR technology advances, utilizing the full potential of non-visual outputs is crucial for enhancing realism and interaction in virtual and augmented environments.
        For instance, delivering content through olfactory sensations enhances the presence and plausibility of immersive XR experiences more effectively than instructing users to "imagine smelling grass" through visual text. However, employing many of these modalities necessitates specialized hardware and software, and their implementation relies on the specifications of the XR system. 
        \par
\begin{figure}[htb]
            \centering
            \includegraphics[width=.45\textwidth]{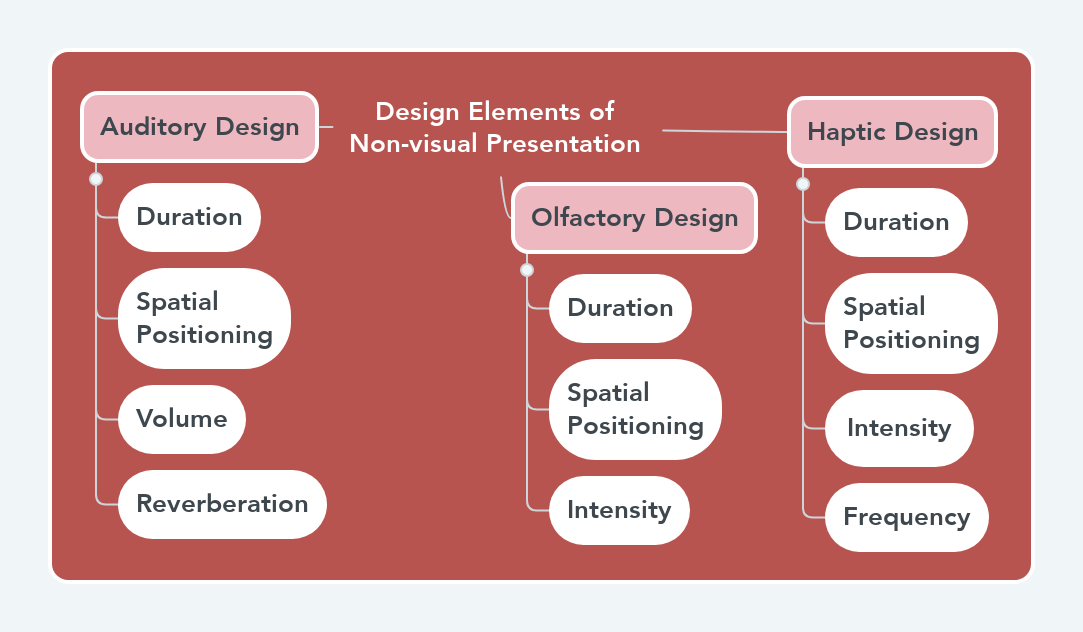}
            \caption{XR design elements for non-visual modalities}
            \label{fig:fignonvisual}
        \end{figure}

        This research primarily concentrates on visual presentation in AR. However, comprehending the design elements of non-visual modalities allows AR designers to develop more immersive, intuitive, interactive, and engaging experiences for users across various applications and contexts. We briefly examine the common design elements for delivering information through non-visual modalities and encourage researchers to further investigate and expand upon this line of inquiry (See \Cref{fig:fignonvisual}).
        \\
        \\
        \textbf{\textit{Audio Presentation}}\\
        Delivering XR object content through audio outputs, such as speech synthesis or sound effects, via the \textit{audio modality} is relatively commonplace. Utilizing speech specifically enables intuitive communication of complex information, reduces ambiguity, and enhances accessibility for visually impaired users.
        We consider the design elements of audio output in XR, focusing on its duration, volume, spatialization, and reverberation.
        The \textbf{duration} of audio output in an XR interface can significantly influence distractions, disturbances, and conflicts.
        Prolonged audio presentations may divert the user's attention away from their task and influence their awareness of the XR environment. Additionally, repetitive or non-verbal audio played for extended periods can become annoying and disruptive.
        Furthermore, aligning the \textbf{spatial positioning} of the audio output in 3D space with its associated XR object and surroundings enhances realism.
        \par
        Matching the audio output's \textbf{volume} and intensity to the XR environment's noise level ensures audibility without disruption. 
        Moreover, the audio output can employ reverb effects to authentically mimic the acoustic properties of the surrounding XR environment. \textbf{Reverberation} artificially creates a spatial perception by imitating how sound bounces off surfaces in different environments, thereby shaping the acoustic ambiance.
        \par \medskip\noindent
        \textbf{\textit{Haptic Presentation}}\\
        The \textit{haptic modality} employs tactile sensations such as vibration, force, and thermal feedback to convey diverse information. For instance, force feedback can signify pressure and weight, while subtle variations can convey texture, shape, and deformation.
        We explore the design elements of haptic output in XR, focusing on its duration, spatial positioning, intensity, and frequency.
        The \textbf{duration} of haptic sensations during XR interaction varies based on context and purpose. Brief vibrations may signal events or interactions, while prolonged feedback can simulate ongoing physical interactions or environmental cues.
        Moreover, the \textbf{spatial positioning} of haptic sensations in the XR environment, relative to the user's body or XR objects, affects the realism and efficacy of haptic output. Aligning haptic output with visual or auditory cues enhances immersion and intuitiveness, enabling users to perceive virtual interactions more accurately.
        \par
        User engagement and immersion in XR can be influenced by the \textbf{intensity} or amplitude of haptic sensations experienced by the user. These sensations simulate a broad spectrum of tactile experiences, ranging from intense vibrations, temperatures, or weights conveying a heightened sense of impact or force to gentle taps.
        Similarly, the \textbf{frequency} of haptic vibrations and output can shape user perception of the content and evoke various sensory experiences. High frequencies may signal urgency, while low frequencies can mimic gentle movements or subtle textures.
        \par \medskip\noindent
        \textbf{Olfactory Presentation}\\
        The \textit{olfactory modality} utilizes scent to convey information or evoke a sense of smell. It reproduces the natural scents of the virtual environment through diffusers, fans, scented materials, electronic devices, and computational methods. 
        Due to technical complexities, such olfactory outputs remain uncommon in XR and are primarily reserved for short, stationary immersive experiences. We discuss the main design elements of olfactory output in XR, focusing on duration, spatial positioning, and intensity.
        Considering the \textbf{duration} of a scent within the design of an XR environment is crucial to ensuring its alignment with user interactions and the experience's narrative.
        On the other hand, strategic \textbf{spatial positioning} of scent sources relative to the user's position within the XR environment can heighten the sense of presence and realism. For instance, positioning floral scent emitters near XR objects depicting flowers enriches the immersive experience by reinforcing visual and olfactory cues.
        Additionally, a scent's \textbf{intensity} / potency, from subtle hint to powerful odor, can evoke various emotional responses or amplify specific narrative elements.
        \par \medskip\noindent
        \textbf{Hybrid Presentation}\\
        The hybrid presentation involves multimodal XR objects that integrate sub-objects of various modalities to present the content.
        Visual and auditory modalities are frequently integrated to enhance the efficacy of an XR object. For example, combining sound effects with visual notifications or background music with visual animations enhances their capacity to convey intended information to the user.
        Similarly, haptic and olfactory modalities are not commonly utilized individually, as the information they convey tends to be ambiguous and limited in range.
        However, they can be effectively employed in conjunction with visual or auditory modalities to enhance presence and plausibility in tailored immersive XR experiences.
        For example, an XR object in an immersive experience can integrate haptic and visual modalities to simulate various temperatures that match the XR object's 3D visual presentation.
        In more general-purpose scenarios, these modalities can also provide cues or signals to the user within a multimodal XR object and increase the efficacy of information presentation.

        \subsection{Input Design} \label{se:inDesignSpace}
            The efficiency and functionality of an XR interface depend on the input design and user interactions as well. Key input design elements of XR encompass the interactivity of each XR object and the selected input modality and interaction design for individual XR objects.
            \par
            The \textbf{interactivity} of an XR object represents users' active engagement to provide input and adapt its design elements, including actions like clicking buttons, entering text, repositioning, or adjusting audio levels.
            An interactive XR interface enhances efficiency by enabling users to customize it, control information flow, and access relevant data according to their needs.
            An XR object can offer an adaptive level of interactivity for specific design elements.
            For instance, typically, the availability of XR objects is interactive, allowing the user to open and close them.
            However, some XR objects, for instance, static visual assets like walls in a VR game, may be designed with no interactivity even in terms of their availability, limiting the user to ``close" and remove them from the environment.
            Additionally, intelligent XR interfaces can adapt the level of interactivity for individual XR objects and particular design elements depending on the context. For instance, they can limit the availability of specific XR objects to restrict students' content access during an exam.
            \par
            Input modality and interaction design encompass the methods used to capture and interpret user input. These are crucial elements in XR interface design influencing ease of use, intuitiveness, accessibility, and efficiency. Inappropriate selection of input modality and poorly crafted interactions can result in user frustration and challenges.\\
            \textbf{Input modalities} range from natural inputs like eye-based, hand-based, and voice-based methods to specialized equipment like controllers and touchpads. The interplay between input modality and user context shapes the XR experience, affecting users' perception of their virtual and physical surroundings and their ability to interact with the interface. Even within a chosen input modality, selecting the right \textbf{interaction technique} significantly impacts users' ability to navigate and utilize the XR interface, influencing its learnability, usability, and overall effectiveness.
\section{Adaptive Environment-referenced Placement} \label{se:caPlacement}
A placement strategy determines the $reference_{position}$ and $reference_{orientation}$ of an XR object's FoR, as well as the pose of the object within its FoR.
The literature has investigated different world-fixed or user-fixed placement strategies for XR objects, including surround-fixed, object-fixed, body-fixed, head-fixed, and device-fixed \cite{bowman2001interaction, feiner1993windows, davari2020occlusion}.
These placement strategies utilize empirical data from previous use cases to either statically pose the XR content within a uniform FoR or propose a context-specific adaptive placement. 
In this work, we propose an adaptive strategy for utilizing each XR object's association with the environment to dynamically place it without prior knowledge of the specific context or environment.
\par
We propose \textit{\textit{Environment-referenced}} placement, an adaptive placement strategy that employs a hybrid FoR tailored to each XR object utilizing an $reference_{orientation}$ distinct from the XR object's $reference_{position}$.
Bell \etal investigated placing relevant objects in close proximity as a strategy for view management \cite{bell2001view}.
This approach intelligently identifies a relevant object or area within the environment as an intermediary for each specific XR object by leveraging machine learning (ML) and large language models (LLMs).
Using each intermediary's relative position to the XR object's $reference_{position}$ our placement strategy determines a distinct $reference_{orientation}$ for each XR object. 
In essence, while the XR content moves along with and remains anchored at a consistent distance from its $reference_{position}$, it continuously rotates to maintain a persistent pose between its $reference_{position}$ and intermediary (See \Cref{fig:ExperimentDiagram}: \textbf{1-b and 2-b}).
\par
Implementing this adaptive placement strategy within a context-aware interface that can identify relevant intermediaries for the user's task and specific context allows for placing the XR objects precisely where needed, thereby enhancing the interface's efficiency.
By leveraging the user's spatial awareness and focusing their attention on nearby RW and virtual objects within their visual range, this placement strategy seamlessly integrates XR objects into the environment, simplifying the interface complexity \cite{krevelen2010survey}.
Furthermore, the intelligent detection of intermediaries within new environments and tracking their relative position to the XR object's $reference_{position}$ offers an adaptive placement strategy for portable and dynamic use cases. 
This is particularly useful when users move to new or unfamiliar environments, or when the position of the user, $reference_{position}$, or intermediaries change within the same environment.
However, in such contexts, especially when the selected intermediaries are irrelevant to the setting and the user's task, this placement strategy can be unintuitive and introduce additional mental and physical load to locate the intermediaries.
\figExperimentDiagram
Envision utilizing various placement strategies for name tags at a conference.
A non-adaptive object-fixed placement strategy might use each attendee as the uniform FoR for their name tag. 
In this interface, the pose of each name tag depends on the specific attendee's position and orientation. This can result in the inconsistent appearance of name tags, creating a confusing interface. For instance, depending on people's relative orientation, one person's name tag might appear to their right while another's name tag appears to their left in the user's view.
Additionally, an attendee's name tag might appear too small or at an angle, depending on their distance from the user or whether they are facing the user, affecting its visibility and readability.
On the other hand, using the user as the uniform FoR ensures consistency in the distance and angle at which name tags appear in the user's view. However, without additional contextual information and adaptations to clearly associate each name tag with the corresponding attendee, a non-adaptive body-fixed placement strategy can result in visual clutter and overwhelm the user with excessive information.
\par
Our placement strategy enables a solution to consider the user as the $reference_{position}$ and each attendee as their intermediary to their name tag.
This enables each person's name tag to appear in a uniform pose around them in the user's view.
Using the user as the $reference_{position}$ ensures consistency in the distance and angle at which the name tags appear in the user's view.
While these XR name tags consistently follow the user, their distinct $reference_{orientation}$ ensures they are visible solely when the specific attendee, i.e., intermediary, is within the user's field of view, minimizing visual clutter and information overload.

\section{Experiment} \label{se:experiment3}
To address \textbf{\textit{RQ3.}}, we designed an experiment to assess information acquisition using our proposed adaptive placement strategy compared to a non-adaptive placement strategy across four distinct contexts: \textit{Dynamic Mobile}, \textit{Dynamic Stationary}, \textit{Static Stationary}, and \textit{Static Mobile}.
\par
This experimental design aims to investigate the nuanced interplay between placement strategy and contextual components.
It is important to note that, while our proposed adaptive strategy intelligently identifies intermediaries, this work does not aim to propose, implement, or evaluate the performance of ML models in identifying optimal intermediaries.
Consequently, to ensure that the reliability and performance of the intermediary detection do not influence our results, within our experimental design we manually identify relevant intermediaries for the XR content within each context.
\subsection{Experimental Design}    
Our experiment incorporated three independent variables each with two levels: 
\textit{Placement Strategy} (adaptive \textit{Environment-referenced} or non-adaptive \textit{Body-fixed}), 
\textit{Real-World (RW) Setting} (either \textit{Dynamic} or \textit{Static}), and
\textit{User State} (\textit{Mobile} or \textit{Stationary}).
We conducted a within-subjects user-study to evaluate the performance of each \textit{Placement Strategy} within each combination of \textit{RW Setting} and \textit{User State}. Thus, each participant's performance was evaluated across eight sessions.
    \subsubsection{Study Task}
    The challenges associated with information acquisition and improper placement of XR content are exacerbated when dealing with a large amount of content. 
    We devised a trivia game about 28 different countries to assess information acquisition within a large collection of AR documents.
    The trivia questions covered three categories: international food, movies, and sports.
    Our AR interface displayed a total of 84 documents, one for each category and country.
    The documents were displayed across three panels in the interface, one for each category.
    The user's objective was to identify the category and country mentioned in the question and open the correct document. 
    All questions explicitly stated both the category name and the country name.
    The question was repeated every 30 seconds until the correct document was opened.   
    \figdocument
    \subsubsection{RW Setting}      
    The user was positioned in an empty room featuring three designated intermediary objects, each representing one of our categories. 
    \par
    \par\medskip\noindent
    \textit{\textit{Static Individual Setting:}}
    In the \textit{Static} contexts, the user participated in a trivia game alone in the room, where questions were displayed on an AR screen in front of them.
    In these \textit{Static} contexts, the RW environment involved three virtual posters, each displaying images relevant to a specific category, and remained static.
    \par
    \par\medskip\noindent
    \textit{\textit{Dynamic Social Setting:}}
    In the \textit{Dynamic} contexts, three virtual humans acted as trivia hosts.
    These trivia hosts engaged in dialogue with the user and each other, moved around the room, and switched positions throughout the session.
    The virtual avatars were each attired in clothing pertinent to one of our categories: a chef, a soccer player, and a director.
    Each host presented questions aligned with their respective category.
    \subsubsection{User State}
    At the beginning of each session, the user was standing at the \textit{Start Point}, facing the sports intermediary.
    \par
    \par\medskip\noindent
    \textit{\textit{Stationary User:}}
    In the \textit{Stationary} contexts, the participant remained consistently positioned at the \textit{Start Point}.
    \par
    \par\medskip\noindent
    \textit{\textit{Mobile User:}}
    In contrast, in the \textit{Dynamic} contexts, participants were prompted to move near a different intermediary at specific times during each session.
    \subsubsection{Placement Strategy}
    We designed a non-adaptive and adaptive placement strategy to display these panels (Refer to \Cref{fig:ExperimentDiagram}). 
    Variations in the spatial layout design elements of the panels influence not only their placement but also their appearance, occlusion, and ease of interaction.
    This emphasizes the importance of consistently presenting the panels at a uniform size, height, and distance within the user's view to ensure that our results are solely due to placement.
    Additionally, to access the panels in \textit{Mobile} contexts, they must follow the user.
    To address these concerns, in both placement strategies, we used the user as the $reference_{position}$ and designed all category panels in a uniform $pose$ and $size$.
\\
\\
\noindent
    \textit{Environment-referenced Placement:}
    As described in \Cref{se:caPlacement}, we implement an adaptive placement strategy to determine an individual $reference_{orientation}$ for each category panel utilizing its intermediary's relative position to the user.
    In our design, we placed each panel directly between the user and its intermediary (See \Cref{fig:socialObj}).
    To ensure that intermediary detection's performance does not influence our results, within each setting, we manually assigned each panel's relevant intermediary.
    In \textit{Static} contexts, the virtual poster, displaying images relevant to each category, served as that category's intermediary.
    \par
    \figsocialObj
    In social interactions, visual engagement with conversation partners is crucial.
    Research by Hood~\etal demonstrates that individuals tend to gaze at their conversation partners for a significant portion of the interaction \cite{hood2021convoSkill}.
    Miller \etal revealed that the mere presence of virtual agents influences the user’s non-verbal and social behavior, with participants avoiding avertion of their gaze from avatars and treating them as real individuals \cite{miller2019social}.
    Consequently, in \textit{Dynamic} contexts, we considered each trivia host as the intermediary for their category panel.
    \figsocialUser
\\
\\
\noindent
    \textit{Body-fixed Placement:}
    As a baseline non-adaptive placement strategy, we implemented a commonly-used \textit{Body-fixed Placement}, in which the user is considered as the uniform FoR.
    By using the orientation of the user's body as the $reference_{orientation}$ for all panels, within this non-adaptive placement, a particular panel is consistently placed on the same side of the user's body. 
    In our implementation, we situated the sports panel directly in front of the user, the food panel to their right side, and the Movies panel to their left (See \Cref{fig:socialUser}).
\subsection{Hypotheses}
Both adaptive and non-adaptive placement strategies have unique advantages that influence their effectiveness for the same task.
Regarding \textit{RQ3}, we hypothesize that neither of the two placement strategies would be singularly superior for information acquisition in all contexts, as their advantages and disadvantages can vary depending on the specific context.
To explore the impact of different context components on the effectiveness of these placement strategies for information acquisition, we formulated and tested three hypotheses denoted as \textbf{H1, H2, and H3}. 
\\
\\
\noindent\textit{\textbf{H1.} Placement strategy effectiveness varies across different contexts.}\par
When utilizing the appropriate contextual information for adaptation, the \textit{Environment-referenced} placement facilitates rapid access to information, by placing each XR object where it is needed.
Within this experiment, this adaptive strategy can lead to intuitive placement and optimal access to information with no added cognitive load for navigation between category panels. 
On the other hand, \textit{Body-fixed} placement facilitates constant and rapid access to all category panels at all times, regardless of the \textit{RW Setting} or \textit{User State}, or the suitability of the intermediaries.
However, this strategy's non-adaptive placement increases the decision-making and mental load associated with navigating the correct category panel.
%
\\
\\
\noindent \textit{\textbf{H2.} The effectiveness of \textit{Environment-referenced} placement strategy is influenced by RW Setting.}\par
The people, objects, events, and elements within the RW setting can affect user behavior, information needs, and interests.
  
We evaluated the performance of adaptive and non-adaptive placement strategies within various \textit{RW Settings}, hypothesizing that \textit{Environment-referenced}:
\par\noindent
\textit{\textbf{H2-a} outperform \textit{Body-fixed} in \textit{Dynamic} contexts.} \\
In social contexts, information acquisitions are usually associated with their conversation and interlocutor with whom the users typically maintain eye contact \cite{hood2021convoSkill}. 
Similarly, in both \textit{Dynamic} contexts within this experiment, the information acquisition inquiries are associated with the same trivia host with whom the user is conversing and visually engaged.
Additionally, the user's engagement in social and conversational interactions, along with the dynamic interactions between the trivia hosts, can increase the user's mental load in our \textit{Dynamic} contexts.
By considering the conversation partner as the intermediary in these contexts, \textit{Environment-referenced} intuitively displays solely the required category panel where needed.
We hypothesize that in these \textit{Dynamic} contexts, the \textit{Environment-referenced} strategy facilitates the cognitive and physical load of finding and accessing the relevant panels and enhances information access.
\par\noindent
\textit{\textbf{H2-b} underperforms \textit{Body-fixed} in \textit{Static} contexts.}\\
The advantages of \textit{Environment-referenced} rely on its potential to place the AR content precisely where needed.
On the other hand, the suboptimal selection of intermediaries can make this placement unintuitive and lead to increased cognitive load and physical effort for information access.
Unlike the \textit{Dynamic} contexts, in our \textit{Static} ones, the user's visual attention within the environment varies, posing challenges in anticipating the optimal intermediaries.
Additionally, due to the unchanging environment and the \textit{solo} social setting, our \textit{Static} contexts impose a lower mental load on the user.
Considering the limited number of category panels, this higher cognitive capacity can facilitate decision-making associated with the \textit{Body-fixed} placement.
We speculate that the \textit{Body-fixed} placement's consistency and effortless access to all documents through a glance, outweigh its cognitive decision-making in \textit{Static} contexts.
Consequently, we hypothesize that in \textit{Static} context, the \textit{Body-fixed} placement's consistent glanceable access enhances information acquisition, compared to \textit{Environment-referenced}. 
\\
\\
\noindent \textit{\textbf{H3.} \textit{Environment-referenced} underperforms \textit{Body-fixed} in \textit{Mobile} contexts.}\par
The user's physical, cognitive, and emotional states influence their capability to interact with and perceive the RW environment, including their perspective, spatial awareness, directional perception, and distance from different objects.
While within our \textit{Stationary} contexts, the user's perception and interaction remain unchanged, in our \textit{Mobile} contexts, users move across different locations within the environment.
  \par
When utilizing \textit{Environment-referenced} placement, depending on the user's position, the relative position of each category panel changes, potentially appearing on their left side at one time and on their right side at another.
The user's varying perspective and directional perception of the environment can lead to inconsistency and confuse the user.
In our \textit{Mobile} contexts, the \textit{Environment-referenced} placement requires spatial recall for the users to determine their current perspective and the relative position of each intermediary, increasing their cognitive load, especially within unfamiliar environments.
Conversely, regardless of the user's mobility and location, \textit{Body-fixed} placement ensures a persistent approach to position each category panel relative to the user.
We evaluate the effectiveness of each placement strategy and hypothesize that the consistency of the non-adaptive \textit{Body-fixed} placement enhances information acquisition in \textit{Mobile} contexts.
\subsection{Apparatus}
Throughout the experiment, participants utilized a Microsoft HoloLens (2nd gen) Augmented Reality Head-Worn Display (AR HWD). This wireless apparatus has a display density of 47 pixels per degree and a diagonal field of view spanning 52 degrees. Our virtual environment and task-specific panels and documents were developed using the Unity 2022.3.4f1 game engine in conjunction with Microsoft's MRTK toolkit.
\par
All sessions started with the user positioned at the \textit{Start Point}, situated at an equal distance from the intermediaries (\Cref{fig:ExperimentDiagram}: \textbf{1-a and 1-b}).
The participant was directly facing the sports intermediary, with the food intermediary to their left, and the movies intermediary to their right. 
We designed eight unique trial sets, each serving as the script for one session. 
Each \textit{Stationary} session comprised three trials, with one trial for each category.
In all \textit{Stationary} trials the user was far (farther than $1.5~m$) from all intermediaries.
In specific \textit{Mobile} trials, the user was positioned near the intermediary relevant to the trial's question.
To ensure the independence of our trial results from this distance, the \textit{Mobile} sessions consisted of six trials, with two trials for each category, one when the user was near its intermediary, and another one when far.
\par
The interface provided three panels, each containing 28 documents corresponding to one of our three trivia categories. These documents were organized alphabetically in a grid format, with four rows and seven columns within each panel. All three panels shared a uniform appearance, except for the panel header, which showcased the category title along with an icon representing that category.
Participants utilized hand interactions facilitated by the HoloLens device to select and interact with documents, enabling them to open and close them as needed. Each document displayed a single word providing the answer to the question, closing once the user clicked anywhere on it. 
        \figquestion
\par
In the \textit{Static} contexts, a screen was displayed in front of the user at eye level to present the trivia question (\Cref{fig:question}).
In the \textit{Dynamic} contexts, a trivia host engaged in a conversation with the user and verbally asked the question.
Consequently, the questions were presented differently across contexts, providing text in the \textit{Static} contexts while audio in the \textit{Dynamic} ones. We employed multiple methods to prevent participants from initiating document searches during the question in specific contexts and ensure consistency across sessions. First, all questions were formulated to reveal the name of the country as the concluding word of the question. Second, in \textit{Dynamic} contexts, participants were instructed to maintain eye contact during questions and to respond only after the entire question was asked.
Third, we matched the rate at which the question was presented across various \textit{RW Settings}.
Using the duration of each question's audio within the \textit{Dynamic} contexts, its equivalent question was revealed one word at a time in the \textit{Static} contexts.
Fourth, while presenting the question, the panel headers were rendered transparent. 
\par
Additionally, within \textit{Dynamic} contexts, to enhance the ecological validity of social interactions and alleviate any biases arising from individual differences and social behavior among participants, we instructed all participants to engage with the trivia hosts as if they were real people and to maintain eye contact during conversations. These trivia hosts were designed using avatars and animations sourced from the Microsoft Rocketbox library and Maximo \cite{gonzalez2020rocketbox, giovannelli2023gestures}.
\subsection{Procedure}
    Following approval from our university’s Institutional Review Board, we enlisted participants from our local university. 
    The experiment lasted 120 minutes and was divided into four parts, each session corresponding to a distinct context. Within each context, each participant underwent two sessions, each employing one of our placement strategies. Overall, each participant underwent eight sessions.
    \par
    The presentation order of the sessions was counterbalanced to mitigate any potential bias arising from a specific trial set or the order in which participants experienced the session.
    To counterbalance the ordering of the sessions, participants were presented with the four contexts in a Latin Square ordering. Within each context ordering, half of the participants began with \textit{Body-fixed} placement, while the other half started with \textit{Environment-referenced}. We recruited 24 subjects, ensuring three participants per session-ordering (3 x 4 x 2).
    \par
    Upon their arrival, participants were asked to carefully review and sign the consent form. Subsequently, we gathered demographic details and information regarding their prior experience with XR. Once this process was completed, we provided a concise overview of the study, acquainted participants with the device, and instructed them on its operation and interaction methods. They then proceeded to engage in the four parts of the experiment.
    \par
    We started each part by introducing the particular task associated with that context. Within each part, participants experienced each \textit{Placement Strategy} in a separate session. Prior to each session, they underwent a training session, completing a minimum of six trials to acquaint themselves with the environment, tasks, and specific content arrangement offered by the \textit{Placement Strategy} within that context.
    Following each session, participants were directed to complete the post-session survey on a computer. 
    During the post-context interview conducted after trying both placement strategies within a context, participants were prompted to compare and rank the two \textit{Placement Strategy} options based on speed, ease of use, intrusiveness, and preference.
    \par
    At the conclusion of the study, all participants underwent interviews regarding their overall experiences across all contexts, as well as the advantages and drawbacks of each \textit{Placement Strategy} within each combination of \textit{RW Setting} and \textit{User State}. Participants had the freedom to take breaks between sessions as needed.

\subsection{Evaluation Metrics}
    The impact of XR on mental workload and task performance is frequently assessed through objective metrics like task completion time and error rate, as well as subjective evaluations like NASA TLX \cite{jeffri2021arEffect}.
    To explore information acquisition using our proposed placement strategies within these contexts, we evaluated the Accuracy, Efficiency, and Usability of each \textit{Placement Strategy} for information acquisition.
    To assess these metrics, we collected objective task-specific data from each trial and subjective data from survey and interview responses.
    \subsubsection{Accuracy}
        A critical aspect in determining the effectiveness of an AR placement strategy for information acquisition is its potential to maintain task performance accuracy.
        In this experiment, a trial continued until the participant opened the correct document.
        We regard every instance of opening the wrong document as an error, adversely affecting accuracy.
        To assess accuracy, we evaluated 
        \textbf{\textit{Error Rate}} 
        based on the total number of trials in which any errors occurred.
    \subsubsection{Efficiency} 
        An effective placement strategy facilitates efficient and swift information acquisition through a large quantity of AR documents and minimizes the associated physical and mental load.
        We assessed information acquisition efficiency through various metrics evaluating the time and difficulty of finding the correct document within each session.
        \par 
        \textbf{\textit{Navigation Time}}: utilized the logged time-stamped gaze data to measure the duration of time for answering a question.
        This time was measured from the moment a question was completely presented to the user for the first time until the user's gaze fixated on the correct document.
        Unfortunately, we observed that the accuracy of the device in recognizing hand gestures was significantly influenced by the participant's race and skin color, resulting in longer times to open a document for certain participants. To remove this bias from our data, we measured \textit{Navigation Time} using the gaze data rather than the time at which a document was opened. 
        \par 
        \textbf{\textit{Gaze Switch}}: measured the frequency of gaze switches among the category panels within each trial.
        An increase in the frequency of gaze switches suggests a higher head-turn rate and physical load for accessing the document.
        Additionally, this measurement can be indicative of the unintuitiveness of the placement and the user's difficulty in navigating the correct category panel.
        \par 
        \textbf{\textit{Perceived Speed}}: quantified the participants' response to the post-context interview question about ranking the fastest placement strategy for information acquisition within that context.
        \par 
        \textbf{\textit{Perceived Ease}}: quantified the participants' post-context ranking of the easiest placement strategy for information acquisition.
        \par 
        \textbf{\textit{NASA TLX}}: collected the post-session response to the short version of the Standard NASA Task Load Index (TLX) questionnaire \cite{nasaTLX1988hart}.
        \subsubsection{Usability}
        To assess the usability of our placement strategies within these contexts:
        \par 
        \textbf{\textit{Perceived Intrusiveness}}: quantified the post-context ranking of the most intrusive placement strategy within each context.
        \par 
        \textbf{\textit{Preferred Placement}}: quantified the ranking of the most-preferred strategy within each context.
        \par 
        \textbf{\textit{SUS}}: collected participants' responses to the short version of the Standard System Usability Survey (SUS) \cite{sus1996brook} in our post-session survey.
    \subsection{Participants}
    We enrolled $24$ participants.
    This population aged between $18$ and $45$ years ($M = 25.62, SD = 5.24$), and included eight female and thirteen graduate students. $13$ participants had little to no experience with XR, and only $5$ had used XR more than ten times.    
\section{User Study Results} \label{se:results}
To test our hypotheses, we conducted a series of analyses on our quantitative data to assess the effects of our three independent variables: \textit{RW Setting}, \textit{User State}, and \textit{Placement Strategy}.
Our data, collected from 24 participants in a within-subjects study comprising eight sessions, represents a sizeable sample size.
Considering our large sample size (n>50), we conducted Kolmogorov-Smirnov (KS) goodness-of-fit tests on our measurements, revealing the non-normal distribution of our data.
\par
We opted to conduct nonparametric statistical analysis using Aligned Rank Transform (ART) \cite{artAnova2011}.
ART enables parametric statistical analyses, such as Repeated Measures Analysis of Variances (RM-ANOVA), for both continuous and ordinal responses, without the assumption of normality or limiting the number of involved factors.
For all metrics, we applied ART to the data and conducted a factorial non-parametric ART RM-ANOVA to test the significant effect of our independent variables.
For all pairwise comparisons on the interactions that showed a significant effect, we applied the ART-C algorithm for multifactor post-hoc contrast tests with Bonferroni adjustments.
We used an $\alpha$ level of $0.05$ in all significance tests.    
    \subsection{Accuracy}
        Our analyses indicate that \textit{Error Rate} is significantly influenced by \textit{RW Setting} X \textit{Placement Strategy} ($F_{(1, 161)} = 5.25 , p \sim 0.02$).
        \figAccuracySetting
        Pairwise comparisons on the effect of \textit{RW Setting} on \textit{Placement Strategy Error Rates} are illustrated in \Cref{fig:accuracy}.
        These results indicate significantly lower \textit{Error Rate} when using \textit{Environment-referenced}, compared to \textit{Body-fixed} in both 
        \textbf{a)} in \textit{Dynamic} contexts ($p \sim 0.01$, supporting \textbf{H2-a}), and 
        \textbf{b)} in \textit{Static} contexts ($p < 0.0001$, rejecting \textbf{H2-b}).
    \subsection{Efficiency}
    Overall analysis of our quantitative and qualitative data on the \textit{Placement Strategy} efficiency within various contexts, points toward a significant influence of the interaction between \textit{RW Setting} and \textit{Placement Strategy}, supporting \textbf{H2}.
\\
\\
\noindent
    \textbf{\textit{Navigation Time}}: 
    \\       
    Information acquisition within correct trials, where no errors occurred, significantly affect \textit{RW Setting} X \textit{Placement Strategy} on 
    \textit{Navigation Time} ($F_{(1, 769.53)} = 22.32,~p < 0.0001$).
    Our pairwise comparison on this interaction indicated that \textit{Environment-referenced} performed 
    \textbf{a)} statistically faster than \textit{Body-fixed} in \textit{Dynamic} contexts ($p < 0.0001$, supporting \textbf{H2-a}).  
    \textbf{b)} significantly faster in \textit{Dynamic} contexts compared to \textit{Static} ones ($p < 0.0001$ supporting \textbf{H1}) (See \Cref{fig:time}).
    \figNavTime
\\
\\
\noindent
    \textbf{\textit{Gaze Switch}}: 
    \\
    \Cref{fig:gazeContext} indicates that \textit{Gaze Switch} fluctuations within correct trials, is significantly affected by 
    \textit{RW Setting} X \textit{User State} X \textit{Placement Strategy} ($F_{(1, 770.32)} = 18.6,~p < 0.0001$), 
    \textit{RW Setting} X \textit{Placement Strategy} ($F_{(1, 770.91)} = 154.24,~p < 0.0001$),
    and \textit{User State} X \textit{Placement Strategy} ($F_{(1, 769.8)} = 56.69,~p < 0.0001$).
    \figGazeContext
    \par
    Pairwise comparison on \textit{RW Setting} X \textit{User State} X \textit{Placement Strategy} revealed that gaze switches between the category panels when using \textit{Environment-referenced} placement were significantly fewer
    \textbf{a)}
    compared to \textit{Body-fixed} in the \textit{Dynamic Mobile} ($p < 0.0001$), \textit{Static Mobile} ($p \sim 0.001$), and \textit{Dynamic Stationary} ($p \sim 0.009$) contexts,
    and \textbf{b)} in the \textit{Dynamic Mobile} context compared to the \textit{Static Mobile} one ($p < 0.0001$).
    Additionally, this analysis indicated that gaze fluctuations when using \textit{Body-fixed} placement were significantly fewer in
    \textbf{c)} the \textit{Dynamic Stationary} context rather than the \textit{Dynamic Mobile} ($p \sim 0.005$),
    \textbf{d)} the \textit{Static Stationary} context rather than the \textit{Static Mobile} ($p \sim 0.035$),
    \textbf{e)} in the \textit{Static Mobile} context compared to the \textit{Dynamic Mobile} ($p \sim 0.006$).
    \par
    These results revealed that compared to \textit{Body-fixed}, the \textit{Environment-referenced} placement reduces gaze switches in \textit{Dynamic} contexts ($p < 0.0001$, supporting \textbf{H2-a}), \textit{Static} onces ($p < 0.0001$, rejecting \textbf{H2-b}), and \textit{Mobile} contexts ($p < 0.0001$, rejecting \textbf{H3}).
    \par \medskip\noindent
    \textbf{\textit{Perceived Speed}}: 
    \\
    Participants' post-context ranking of the fastest placement strategy revealed a significant interaction effect of \textit{RW Setting} X \textit{Placement Strategy} 
    ($F_{(1, 161)} = 135.01,~p < 0.0001$). 
    Pairwise comparison on the effect of these interactions  
    revealed that \textit{Environment-referenced} was ranked \textbf{a)} faster than \textit{Body-fixed} in the \textit{Dynamic} contexts by 86\% (supporting \textbf{H2-a}),
    \textbf{b)} slower than \textit{Body-fixed} in the \textit{Static} contexts by 84\% ($p < 0.0001$) (supporting \textbf{H2-b}), and
    \textbf{c)} faster in the \textit{Dynamic} contexts compared \textit{Static} ones ($p < 0.0001$, supporting \textbf{H1}).
    Additionally, \textbf{d)} \textit{Body-fixed} was ranked higher in \textit{Static} contexts compared to \textit{Dynamic} ones ($p < 0.0001$, supporting \textbf{H1}).
     
        \par \medskip\noindent
        \textbf{\textit{Perceived Ease}}: 
        \\
        Participants' post-context ranking of the easiest placement strategies revealed a significant interaction effect of 
        \textit{RW Setting} X \textit{Placement Strategy} ($F_{(1, 161)} = 231.4,~p < 0.0001$).
        \textit{Environment-referenced} was ranked \textbf{a)} easier than \textit{Body-fixed} in the \textit{Dynamic} contexts by 93\% (supporting \textbf{H2-a}),
        \textbf{b)} less easy than \textit{Body-fixed} in the \textit{Static} contexts by 80\% ($p < 0.0001$) (supporting \textbf{H2-b}). 
        Indicating that 
        \textbf{c)} \textit{Environment-referenced} placement was perceived as easier in \textit{Dynamic} contexts compared to \textit{Static} ones ($p < 0.0001$, supporting \textbf{H1}), while \textbf{d)} \textit{Body-fixed} placement was periceved significantly more difficult in \textit{Static} contexts compared to \textit{Dynamic} ones ($p < 0.0001$, supporting \textbf{H1}).
        \par \medskip\noindent
        \textbf{\textit{NASA TLX}}: 
        \\
        Our pairwise comparison on the effect of \textit{RW Setting} X \textit{Placement Strategy} on the \textbf{TLX Score}($F_{(1, 161.06)} = 11.68,~p < 0.001$) indicates significantly lower overall task loads,
        \textbf{a)} in \textit{Dynamic} contexts, when using \textit{Environment-referenced} compares to \textit{Body-fixed} ($p \sim 0.001$, supporting \textbf{H2-a}), and 
        \textbf{b)} when utilizing \textit{Body-fixed} in \textit{Static} contexts rather than \textit{Dynamic} ones ($p\sim 0.002$, supporting \textbf{H2-b}).
        \par
        Analyzing the metrics from within the TLX questionnaire identified that 
        \textit{Effort} significantly varied dependent on the 
        specific \textit{RW Setting} X \textit{User State} X \textit{Placement Strategy} ($F_{(1, 161.08)} = 4.51,~p \sim 0.03$, providing support for \textbf{H1 }and\textbf{ H2}), and 
        \textit{RW Setting} X \textit{Placement Strategy} significantly impact 
        \textit{Physical Demand} ($F_{(1, 161.08)} = 6.25,~p \sim 0.01$),
        \textit{Mental Demand} ($F_{(1, 161.10)} = 6.24,~p \sim 0.01$), and
        \textit{Effort} ($F_{(1, 161.08)} = 10.4,~p \sim 0.001$).
        Pairwise comparison on these metrics shows that:
        \par
        Significantly less \textbf{Effort} was required  
        \textbf{a)} in \textit{Dynamic} contexts, when using \textit{Environment-referenced} compared to \textit{Body-fixed} ($p < 0.001$, supporting \textbf{H2-a}),
        \textbf{b)} when utilizing \textit{Body-fixed} in \textit{Dynamic} contexts rather than \textit{Static} ones ($p \sim 0.04$, supporting \textbf{H1}), and 
        \textbf{c)} when using \textit{Environment-referenced} in \textit{Static Stationary} contexts, rather than \textit{Static Mobile} ones ($p \sim 0.02$, providing partial support for \textbf{H3}).
        \par
        \textbf{Mental Demand} is significantly decreased,
        \textbf{a)} in \textit{Dynamic} contexts, when using \textit{Environment-referenced} compared to \textit{Body-fixed} ($p \sim 0.001$, supporting \textbf{H2-a}), and
        \textbf{b)} when using \textit{Body-fixed} in \textit{Static} contexts rather than \textit{Dynamic} ones ($p \sim 0.01$, supporting \textbf{H1}).
        \par
        \textbf{Physical Demand} is significantly lower,
        \textbf{a)} in \textit{Static} contexts, when using \textit{Body-fixed} compared to \textit{Environment-referenced} ($p \sim 0.02$, supporting \textbf{H2-b}), and 
        \textbf{b)} when utilizing \textit{Body-fixed} in \textit{Static} contexts rather than \textit{Dynamic} ones ($p \sim 0.01$, supporting \textbf{H1}).
\subsection{Usability}
Our Analysis indicates a significant interaction effect of 
\textit{RW Setting} X \textit{Placement Strategy} for 
\textit{Perceived Intrusiveness} ($F_{(1, 161)} = 34.82,~p < 0.0001$),
\textit{Preferred Placement} ($F_{(1, 161)} = 246.45,~p < 0.0001$), and 
\textit{Overall SUS Scores} ($F_{(1, 161.05)} = 11.6,~p < 0.001$).
    
    \par \medskip\noindent
    \textbf{\textit{Perceived Intrusiveness}}: 
    \\
    \textit{RW Setting} X \textit{Placement Strategy} significantly influnced  
    \textit{Perceived Intrusiveness} ($F_{(1, 161)} = 34.82,~p < 0.0001$).
    The majority of our participants
    (79.2\% within the \textit{Static} contexts, and 56.3\% within the \textit{Dynamic} ones), insisted that both placement strategies were non-intrusive and refused to choose any one as the most intrusive.
    In \textit{Static} contexts, only 6.3\% ranked \textit{Body-fixed} as intrusive.
    On the contrary, only 4.2\% ranked \textit{Environment-referenced} as intrusive for \textit{Dynamic} contexts.
    Our pairwise comparison indicates that
    \textbf{a)} in \textit{Dynamic} contexts, \textit{Environment-referenced} is significantly less intrusive than \textit{Body-fixed} ($p < 0.0001$, supporting \textbf{H2-a}),
    \textbf{b)} \textit{Environment-referenced} is significantly less intrusive in \textit{Dynamic} contexts compared to the \textit{Static} ones ($p < 0.0001$, supporting \textbf{H1}), and 
    \textbf{c)} \textit{Body-fixed} is significantly less intrusive in \textit{Static} contexts compared to the \textit{Dynamic} ones ($p < 0.0001$, supporting \textbf{H2-b}).
    \par \medskip\noindent
    \textbf{\textit{Preferred Placement}}: 
    \\
    ART RM-ANOVA results on \textit{Perceived Intrusiveness} indicate a significant interaction effect of \textit{RW Setting} X \textit{Placement Strategy} ($F_{(1, 161)} = 246.45,~p < 0.0001$).
    90\% of our participants ranked \textit{Environment-referenced} as the most preferred placement strategy for information acquisition in \textit{Dynamic} contexts, and 82\% ranked \textit{Body-fixed} as the most preferred for \textit{Static} contexts.
    Our pairwise comparison indicates that
    \textbf{a)} in \textit{Dynamic} contexts, significantly more participants preferred \textit{Environment-referenced} over \textit{Body-fixed} ($p < 0.0001$, supporting \textbf{H2-a}),
    \textbf{b)} in \textit{Static} contexts, significantly more participants preferred \textit{Body-fixed} over \textit{Environment-referenced} ($p < 0.0001$, supporting \textbf{H2-b}),
    \figPreferred
    \\
    \textbf{c)} \textit{Environment-referenced} is significantly more preferred for \textit{Dynamic} contexts compared to \textit{Static} ones ($p < 0.0001$, supporting \textbf{H1}), and 
    \textbf{d)} using \textit{Body-fixed} is significantly more preferred for \textit{Static} contexts compared to \textit{Dynamic} ones ($p < 0.0001$, supporting \textbf{H1}). 
    \par \medskip\noindent
    \textbf{\textit{SUS}}: 
    \\
    Our Analysis of the SUS responses indicate 
    significant effect of \textit{RW Setting} X \textit{Placement Strategy} on the
    \textbf{overall Usability Score} ($F_{(1, 161.05)} = 11.6,~p < 0.001$).
    Delving deeper into the metrics from within SUS, found that other than the \textit{overall Usability Score}, more specifically, 
    \textit{User Satisfaction} ($F_{(1, 161.12)} = 16.37,~p < 0.0001$),
    \textit{Synergy} ($F_{(1, 161.1)} = 8.62,~p \sim 0.003$),
    \textit{Consistency} ($F_{(1, 161.07)} = 7.4,~p \sim 0.007$),
    \textit{Learnability} ($F_{(1, 161.09)} = 3.62,~p \sim 0.05$), and
    \textit{Effortlessness} ($F_{(1, 161.13)} = 5.55,~p \sim 0.02$)
    were significantly affected by \textit{RW Setting} X \textit{Placement Strategy}.
    Pairwise comparisons on these measurements provide support for \textbf{H2-a}, indicating that in \textit{Dynamic} contexts, \textit{Environment-referenced} significantly improves overall usability score ($p < 0.0001$) and provides a significantly more satisfactory ($p < 0.0001$), well-integrated ($p < 0.0001$), and consistent ($p \sim 0.04$) interface that better facilitates learning ($p < 0.0001$) and effortless document navigation ($p \sim 0.004$) compared to \textit{Body-fixed}.
    These analyses also support \textbf{H1} by showing that
    \textbf{a)} utilizing \textit{Environment-referenced} resulted in significantly higher user satisfaction in \textit{Dynamic} contexts compared to the \textit{Static} ones ($p\sim 0.03$), and
    \textbf{b)} utilizing \textit{Body-fixed} results in significantly higher user satisfaction in \textit{Static} contexts compared to the \textit{Dynamic} ones ($p\sim 0.005$).
\section{Discussion}
In this section, we evaluate the user study results, examining their alignment with our hypotheses. We explore alternative hypotheses to explain any discrepancies between our initial hypotheses and the findings, offering insights into the study’s limitations and proposing directions for future research.
\subsection{Evaluation of Our Hypotheses}
Our results revealed that the effectiveness of \textit{Environment-referenced} varies across different contexts, excelling in certain scenarios while falling short compared to \textit{Body-fixed} in others.
For each of our measurements, the factorial analysis revealed at least one significant interaction effect involving \textit{Placement Strategy} and one or both of our contextual components, i.e., \textit{RW Setting} and \textit{User State}.  
Considering \textit{Environment-referenced} placement, the participants switched their gaze between panels less often, navigated the documents faster, and expressed overall better feedback on the usability and efficiency of the placement strategy, when in \textit{Dynamic} contexts, compared to \textit{Static} ones.
\textit{Body-fixed} placement increased the user's gaze switches within the \textit{Dynamic} contexts compared to the \textit{Static} contexts. Additionally, considering the \textit{User State}, when \textit{Mobile}, this placement increased gaze switches within a \textit{Dynamic} setting but reduced it within the \textit{Static} setting.
This underscores the impact of contextual scenario on the efficacy of placement strategy, thereby supporting \textbf{H1}.
\par
To evaluate \textbf{H2}, we investigated the performance of \textit{Environment-referenced} information acquisition, 
compared to \textit{Body-fixed} in our \textit{Dynamic} (\textbf{H2-a}) and \textit{Static} (\textbf{H2-b}) contexts.
All of our measurements consistently support \textbf{H2-a}, indicating that in \textit{Dynamic} contexts \textit{Environment-referenced} offers a similar level of physical efficiency as \textit{Body-fixed}, with significantly lower cognitive load, optimizing accuracy, speed, ease, and usability of information acquisition.
This underscores the importance of context awareness, suggesting that in contexts where the relevant intermediary objects are dynamic, applying \textit{Environment-referenced} placement in a context-aware interface could enhance the user experience.
\par
Our data yielded inconclusive outcomes regarding \textbf{H2-b} and \textbf{H3}.
While participants’ post-context ranking of the placement strategies and the Overall TLX score support \textbf{H2-b}, indicating that users perceive \textit{Body-fixed} to outperform \textit{Environment-referenced} in \textit{Static} contexts, the \textit{Error Rate} and \textit{Gaze Switch} measures contradict this hypothesis, indicating lower accuracy and efficiency of \textit{Body-fixed} placement.
Similarly, according to our results, the \textit{User State} only influences placement strategy performance in terms of \textit{Gaze Switch} and the \textit{Effort} subscale of the TLX. Both these metrics indicate the inferior performance of \textit{Body-fixed} in the \textit{Mobile} contexts and refute \textbf{H3}.
\subsection{Influence of Info-focus Relevance}
Our results strongly confirm \textbf{H1} and \textbf{H2-a} while offering partial support for \textbf{H2-b} and partially rejecting \textbf{H3}.
The inconsistency in our results can be attributed to a subtle interplay among our contextual components, which influences the suitability of the intermediaries.
Across these contexts and even within specific ones, the user's focus on specific objects within the room varied.
We used the user's primary conversation partner or location within each trial to extract the user's primary focus.
While the choice of intermediaries ensures that the information access needs are relevant to the intermediaries, within each trial, the relevance of the information being accessed to the user's primary focus differs.
This discrepancy impacted the efficacy of the \textit{Environment-referenced} placement strategy placement strategy across diverse contexts.
We use the term \textit{Info-focus Relevance} to refer to the degree of relevance between information access need and the user's current focus within each trial. 
The following discussion delves deeper into \textit{Info-focus Relevance}'s effect on placement strategy performance in our contexts.
\par \medskip \noindent
\textbf{\textit{Dynamic Contexts}:}
Within our \textit{Dynamic} contexts, the user was always focused on their conversation partner. Additionally, in these contexts, regardless of the \textit{User State}, the information needs within a trial were always \textit{Relevant} to the same trivia host the user was interacting with.
Due to the high \textit{Info-focus Relevance}, the intermediaries are suitable for all trials in the \textit{Dynamic} contexts.
This enables \textit{Environment-referenced} placement to always provide effortless access to the required document through a forward glance and eliminate the cognitive load associated with navigating to the correct category panel.
This can explain our results regarding \textbf{H3}, as the \textit{Environment-referenced} performs significantly better than \textit{Body-fixed} in \textit{Dynamic} contexts, even when the \textit{User State} was \textit{Mobile}.
\par \medskip \noindent
\textbf{\textit{Static Contexts}:}
In the \textit{Static} contexts, the user's focus and, therefore, \textit{Info-focus Relevance} vary based on the \textit{User State}.
The \textit{Static Mobile} context involved the user moving to one of the intermediary objects and maintaining visual attention to it.
In this context, half of the trials require information relevant to the same category as the nearby intermediary object.
Consequently, only half of the trials in this context were \textit{Relevant} in terms of \textit{Info-focus Relevance}. 
\figstaticGaze
On the other hand, within all the \textit{Static Stationary} trials, the user was facing forward and focused on the Sports category's poster.
In this context, only $1/3$ of the trials require information relevant to the Sports category and were considered as \textit{Relevant} in terms of \textit{Info-focus Relevance}.
\par
To better understand \textit{Info-focus Relevance}, we conducted a factorial non-parametric ART RM-ANOVA to test the interaction effects of \textit{User State}, \textit{Placement Strategy} and \textit{Info-focus Relevance} on the quantitative trial data within the Static contexts.
The results indicate significant influence of \textit{Placement Strategy} and \textit{Info-focus Relevance} on \textit{Gaze Switch} ($F_{1, 372.59} = 40.02, p < 0.0001$) (See \Cref{fig:gazeStatic}) and significant effect of \textit{User State} X \textit{Placement Strategy} X \textit{Info-focus Relevance} on \textit{Navigation Time} ($F_{1, 372.39} = 11.65, p < 0.001$) (See \Cref{fig:timeStatic}).
\figstaticTime

\par
\textbf{\textit{Static Stationary}:}
The \textit{Static Stationary} context entails a stationary user positioned between the three posters (intermediaries), maintaining a consistent orientation and location. 
Consequently, \textit{Environment-referenced} appears almost identical to \textit{Body-fixed} and imposes equivalent physical and mental load levels. 
This is contradictory to our \textbf{H2-b} and partially explains the lack of support for this hypothesis in our findings.
Additional nonparametric analyses using Wilcoxon Rank Sums tests 
on our objective data, NASA TLX, and SUS measurements within this context indicate no significant effect of \textit{Placement Strategy} on any of our metrics. 
\par
\textbf{\textit{Static Mobile}:}
We conducted nonparametric analyses using Wilcoxon Rank Sums tests to examine the effect of \textit{Placement Strategy} within this context.
The results indicated that \textit{Environment-referenced} placement in this context resulted in significantly lower \textit{Gaze Switch} compared to \textit{Body-fixed} ($\chi^{2} = 11.84, Z = 3.44, p \sim 0.0006$).
This helps further explain the contradictory results to \textbf{H2-b} and \textbf{H3}.
\section{Limitations \& Future Work}
    In each combination of our context components (\textit{RW Setting} and \textit{User State}), the user’s focus within the environment varied. This resulted in the involvement of \textit{Info-focus Relevance} as an independent variable, influencing the effectiveness of \textit{Environment-referenced} placement strategy.
    On the other hand, \textit{Body-fixed} offers reliable and effortless access to all documents through a glance, irrespective of the particular context and \textit{Info-focus Relevance}. While this placement strategy imposes an additional mental load to select the required information, this mental load is consistent across different contexts. Additionally, as mentioned by our participants, due to the limited number of category panels within this study, this mental load is relatively easy.
    \par
    Our results indicate that contingent upon the user's focus and interactions with their setting, the choice of optimal placement strategy within each context varies.
    In a context-switching scenario, it is essential for a context-aware interface to evaluate the cognitive and physical load associated with various placement strategies and determine the optimal placement for the specific context.
    Such intelligent interfaces must infer information such as \textit{Info-focus Relevance} from the context. In addition, future work must explore various ways for such interfaces to provide feedback to users when the placement strategy changes.
\section{Conclusions}
Towards context-aware adaptations in XR, we explored various XR design elements that can be adapted, providing a comprehensive design space for XR (\Cref{se:designSpace}).
Concentrating on the content’s spatial layout from our design space, we proposed utilizing a relevant intermediary from the environment within a Hybrid Frame of Reference (FoR) for each XR object (\Cref{se:caPlacement}).
We investigated the effectiveness of this adaptive placement strategy and a non-adaptive \textit{Body-Fixed} placement strategy in four contextual scenarios varying in \textit{RW Setting} and \textit{User State} (\Cref{se:experiment3}).
\par
Within our experiment, the \textit{Environment-referenced} strategy utilized relevant people and objects from the environment as intermediaries to position the XR content.
On the other hand, the \textit{Body-fixed} strategy provided a consistent glanceable interface, offering effortless access to all content, regardless of the specific context. 
Our results indicate the significance of context in the choice of optimal placement strategy and highlight the significance of \textit{Info-focus Relevance} on the effectiveness of \textit{Environment-referenced} placement strategy across different contexts. 
In contexts with more \textit{Relevant Info-focus} trials, \textit{Environment-referenced} intelligently positioned each piece of AR content precisely where needed, thereby enhancing efficiency while mitigating information overload.
However, with \textit{Irrelevant Info-focus}, \textit{Environment-referenced} diminished performance as it necessitated excessive head and body movements towards the intermediaries and was inconsistent when the user or the environment was dynamic.
These observations indicate the significance of \textit{RW Setting} and \textit{User State} in determining \textit{Info-focus Relevance}.
More specifically, within our contexts, the interlocutor, the user's proximity to the intermediary objects, and the mobility of the user influenced the user's focus and information needs. Our findings underscore the significance of context-aware interfaces that determine the optimal \textit{Placement Strategy} through inferring information such as \textit{Info-focus Relevance} from the context. 
\acknowledgments{%
Special thanks to our study participants and our colleagues at the 3DI Lab and Adobe for their contributions in shaping the direction of this paper. AI tools were used only in the final round of polishing of this paper.
}
\bibliographystyle{files/bib/abbrv-doi-hyperref-narrow}
\bibliography{bibliography.bib}
\begin{tabular}{ m{0.2\textwidth} m{0.75\textwidth} }
    \begin{minipage}{0.2\textwidth}
        \includegraphics[width=\linewidth]{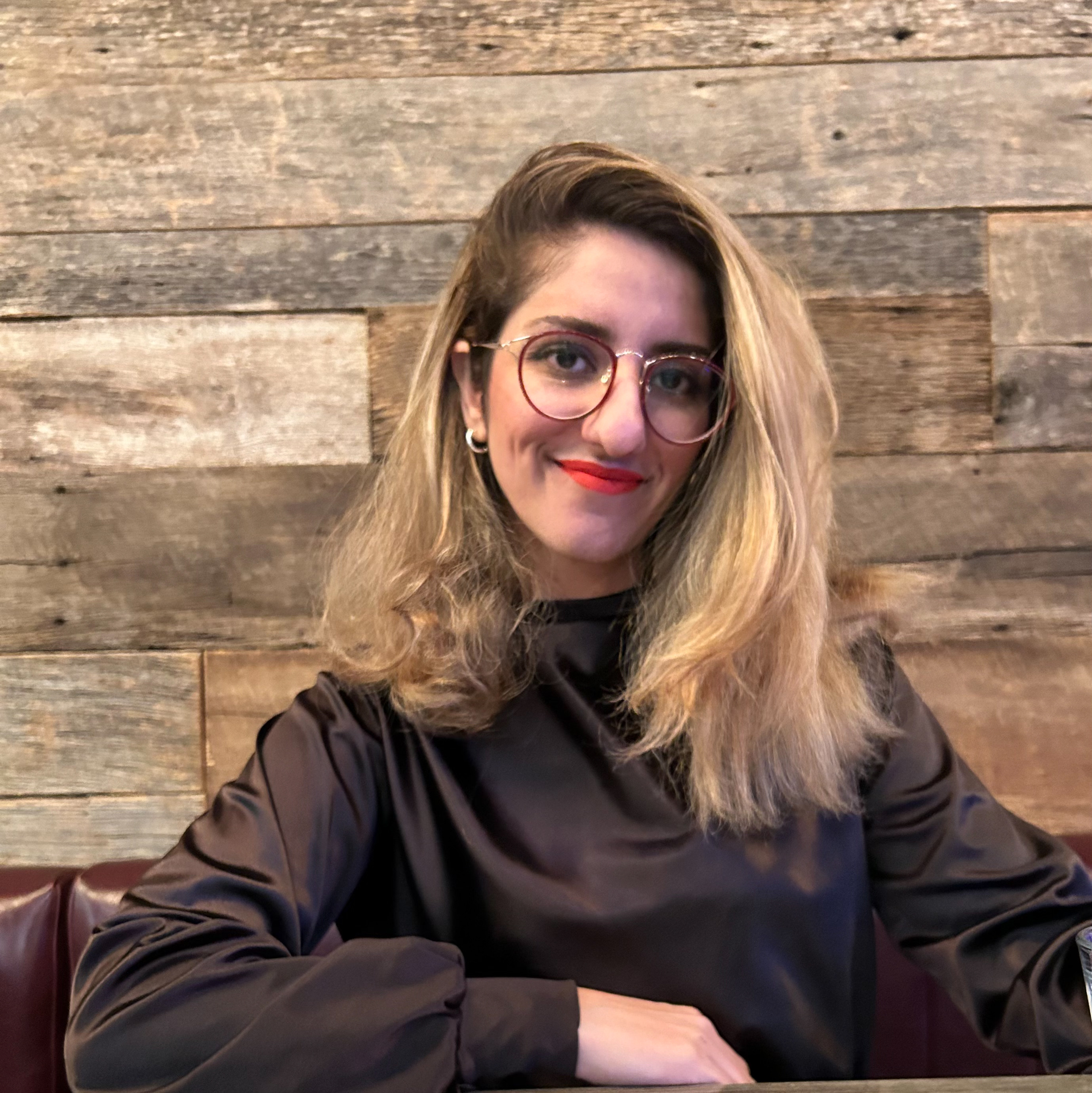}
    \end{minipage} &
    \begin{minipage}{0.27\textwidth}
        \textbf{Shakiba Davari} \newline Shakiba Davari received her Ph.D. from Virginia Tech with a focus on intelligent XR interfaces. Throughout her Ph.D., she has contributed methodological guidelines and frameworks for developing intelligent XR and investigated context-aware AR interfaces that address AR challenges such as real-world occlusion and social intrusiveness.\\
    \end{minipage} \\
    \begin{minipage}{0.2\textwidth}
        \includegraphics[width=\linewidth]{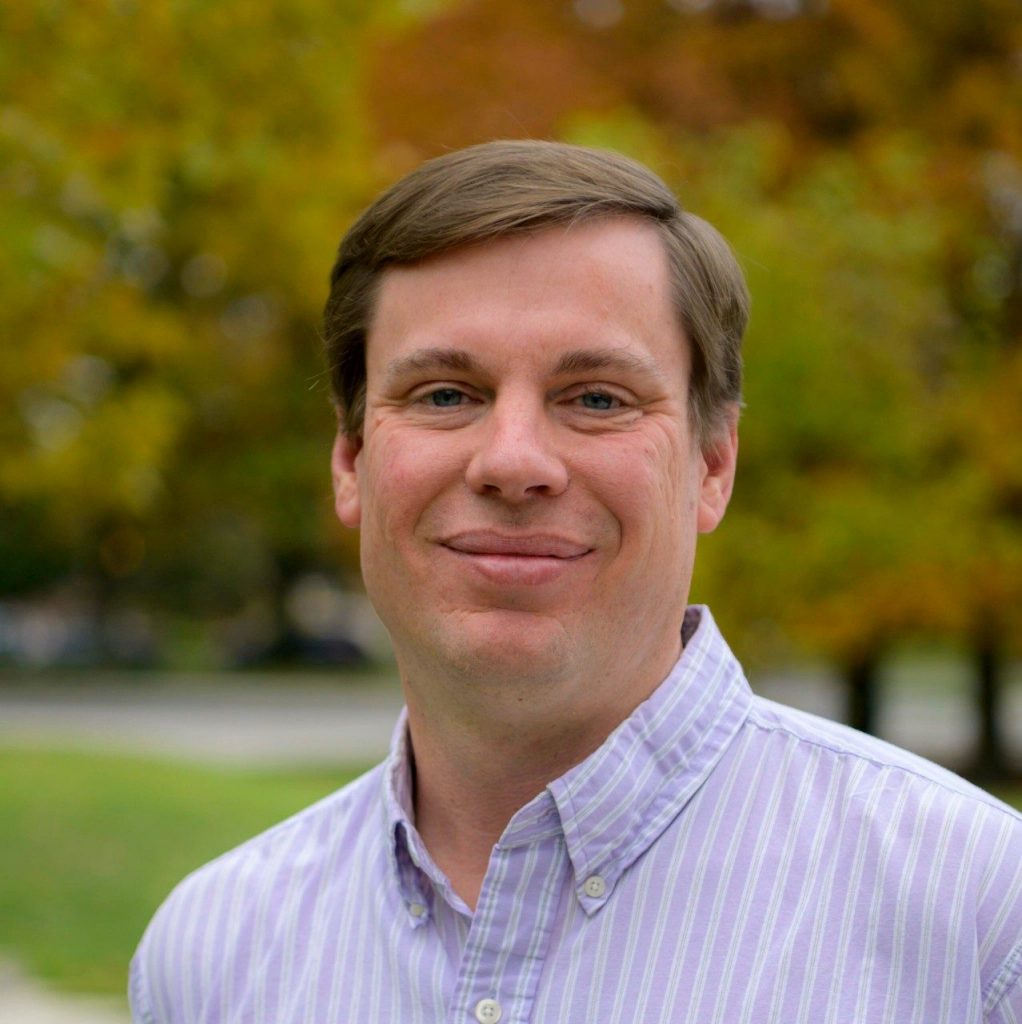}
    \end{minipage} &
    \begin{minipage}{0.27\textwidth}
        \textbf{Doug A. Bowman} \newline Doug A. Bowman is the Frank J. Maher Professor of Computer Science and Director of the Center for Human-Computer Interaction at Virginia Tech. He is the principal investigator of the 3D Interaction Group, focusing on the topics of three-dimensional user interfaces, VR/AR user experience, and the benefits of immersion in virtual environments \\
    \end{minipage}
\end{tabular}
\end{document}